\documentclass[journal]{vgtc}              

\DeclareGraphicsExtensions{.png,PNG,.pdf,.PDF}

\onlineid{1313}

\vgtccategory{Research}

\def \sys {{CELLens}}

\def \etal {{\emph{et al}.\thinspace}}

\def \eg {{\emph{e.g}.}}
\def \ie {{\emph{i.e}.}}

\title{Human-Guided Causal Knowledge Injection for Virtual Cells}

\author{%
  Pengcheng Wang, Changjian Chen, Zhuo Tang, You Wu, Long Wang, Feng Yu and Kenli Li
}

\authorfooter{
  \item
    All the authors are with Hunan University. P.~Wang, C.~Chen and K.~Li are also with Yuelushan Laboratory.
    \\E-mail: \{wangpengcheng, changjianchen, ztang, wuyouray, wanglong8591, feng\_yu, lkl\}@hnu.edu.cn. P.~Wang and C.~Chen are joint first authors. Z.~Tang is the corresponding author.
}

\abstract{Virtual cells employ machine learning models to simulate and predict cellular behaviors, serving as a critical computational framework for investigating health and disease. 
Injecting causal graphs into virtual cells can improve the interpretability, but such graphs are usually not available in real-world applications.
Recently, many methods have been proposed to construct causal graphs from data, which \pcheng{group genes based on their similarities} to form concepts and extract their causal relationships.
However, since this automatic process is \pcheng{unsupervised}, the causal graphs usually contain errors.
In this paper, we propose a human-guided causal knowledge injection method for virtual cells.
We developed a gene-similarity-aware causal graph visualization supported by a hybrid optimization algorithm to help explore both the causal relationships between concepts and the similarities between genes.
Based on the exploration, we further developed a counterfactual analysis strategy supported by a counterfactual visualization and a causal path visualization to help validate and refine causal graphs. 
The effectiveness of our method is demonstrated through two real-world case studies, the extraction of scientifically meaningful causal insights, and positive feedback from domain experts.

}

\keywords{Virtual cell, causal knowledge injection, projection}

\teaser{
  \centering
  \includegraphics[width=1.0\linewidth]{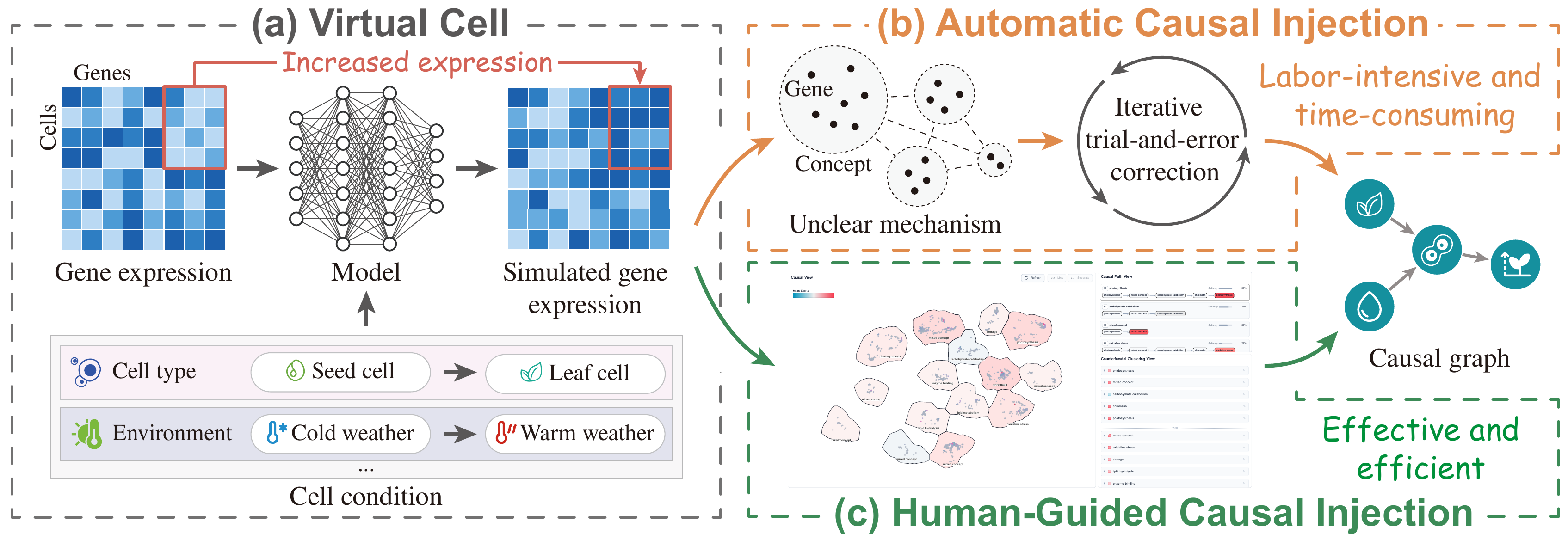}
  \caption{
\pcheng{Overview of virtual cells and the comparison between the automatic and the human-guided causal injection.
(a) A virtual cell simulates cellular behaviors under cell condition changes, such as changes in cell type and environment, and predicts the corresponding gene expression changes.
(b) In automatic causal injection, auto-mined concepts and causal relationships can be unclear or inaccurate, leading to labor-intensive and time-consuming trial-and-error correction before obtaining a reliable causal graph.
(c) {\sys} supports human-guided causal injection by providing an interactive visual interface for effectively and efficiently explore, validate, and refine the causal relationships.}
  }
  \label{fig:teaser}
}

\graphicspath{{figs/}{figures/}{pictures/}{images/}{./}} 

\usepackage{multirow}
\usepackage{color}
\usepackage{lipsum}                    
\usepackage{amssymb,amsmath}
\usepackage{bbm} 
\usepackage{tabularx} 
\usepackage{wrapfig}
\usepackage{booktabs}
\usepackage{makecell}
\usepackage{graphicx}
\usepackage{dblfloatfix}

\newcommand{\myparagraph}[1]{\vspace{1mm}\noindent\textbf{#1}}

\newcommand{\pcheng}[1]{\textcolor{black}{#1}}

\usepackage{mathptmx}                  

\begin{document}


\firstsection{Introduction}

\maketitle
\fontsize{9}{9} 
The \textit{cell} is the basic structural and functional unit of all known forms of life or organisms~\cite{mazzarello1999unifying}.
\pcheng{Understanding cellular behaviors and functions is therefore essential for investigations into health and disease~\cite{regev2017human, rood2025human}.}
However, understanding cells is non-trivial because each cell is a dynamic and adaptive system whose complex behaviors arise from numerous molecular interactions~\cite{bunne2024build}.
To this end, the concept \textit{virtual cell} is consequently proposed.
\pcheng{A virtual cell refers to a machine learning model that simulates cellular behaviors upon cell condition changes, such as cell types and environments.}
\pcheng{Such a computational model offers a powerful and efficient alternative to traditional experimental methods for decoding cellular complexity~\cite{theodoris2023transfer, bunne2024build}.}
\pcheng{For example, in Fig.~\ref{fig:teaser}(a), after changing the cell type and environment, the virtual cell simulates the cellular behaviors and shows increased expression of certain genes, suggesting their potential association with such cell conditions without time-consuming experiments.}

Although virtual cells largely enhance the understanding of cells, recent studies reveal that they often overlook causal relationships between biological factors, \pcheng{which may oversimplify complex biological interactions and thus limit cellular mechanistic understanding~\cite{gao2025causal}.}
Therefore, many causal-driven virtual cell methods have been proposed recently, such as GEARS~\cite{roohani2024predicting} and CausCell~\cite{gao2025causal}.
These methods incorporate causal graphs \pcheng{to guide the virtual cell modeling.}
\pcheng{A causal graph refers to a directed acyclic graph whose nodes represent high-level biological concepts (cell type, batch effects of cells, etc.), and whose edges represent causal influences between these concepts.}
\pcheng{By encouraging the virtual cells to be consistent with the causal relationships defined in the graph, these methods effectively improve the interpretability.}
However, such methods assume that causal graphs are accurate and complete.
In real-world scenarios, causal graphs are usually provided by experts or mined from data automatically, \pcheng{while the former are often incomplete and the latter are often inaccurate}, which severely restricts their broader practical application in virtual cells.

\pcheng{Despite these limitations,} we observed that the accuracy of expert-provided and the completeness of auto-mining methods complement each other~\cite{ruscone2025neko}, \pcheng{motivating us to inject expert knowledge into auto-mined causal graphs.}
However, such a combination is non-trivial due to two main technical challenges:
1) \textbf{Causal \pcheng{graph} exploration and analysis}.
To obtain causal graphs for virtual cells, auto-mined methods typically group similar genes based on their similarities to form concepts and extract their causal relationships using causal discovery methods (\eg, SCENIC~\cite{aibar2017scenic} and CWGCNA~\cite{liu2024cwgcna}).
\pcheng{However, due to noisy expression patterns and the uncertainty of causal discovery, the grouped concepts may contain semantically inconsistent genes, and their causal relationships may include incorrect links or miss important ones.}
Therefore, it is necessary to help experts explore the similarities between genes to understand how they form concepts and explore the causal relationships to verify their correctness.
2) \textbf{Causal \pcheng{relationship} validation and refinement}. 
Once potential incorrect \pcheng{or missing} \pcheng{concept} causal relationships are identified, experts need to validate their impact on virtual cells and refine them if necessary.
However, this process is based on trial and error and heavily depends on the expertise of experts~\cite{hoel2013quantifying}, which makes it labor-intensive and time-consuming (Fig.~\ref{fig:teaser}(b)).
Therefore, a visual tool that facilitates efficient causal validation and refinement is desirable.

To address these challenges, we develop {\sys} (Fig.~\ref{fig:teaser}(c)), a visual analysis \pcheng{tool} to help 1) efficiently explore causal relationships and 2) validate and refine them to ensure biological reliability, \pcheng{following an expert-centered design-study process that grounds the biological requirements, design decisions, and evaluation in realistic analysis scenarios.}
Given a set of cells with their gene expression data, a causal graph is constructed by grouping genes into concepts and extracting causal relationships between these concepts.
Compared to normal causal graphs \pcheng{that only contain concept-concept causal relationships}, this causal graph additionally includes the gene-concept correspondences and similarities between genes.
While many causal graph visualization methods (\eg, CausalVis~\cite{guo2023causalvis}) have been proposed in the literature, they primarily focus on preserving concept-concept causal relationships and fail to preserve concept-gene correspondences and gene-gene similarities.
To this end, we propose a gene-similarity-aware causal graph visualization supported by a hybrid optimization algorithm, \pcheng{to present concept-level causal relationships along the horizontal direction while preserving gene similarities within each associated concept.}
\pcheng{With this visualization,} experts explore the causal graph and identify the potential incorrect \pcheng{or missing} causal relationships.
To validate these potential incorrect \pcheng{or missing} ones, \pcheng{we further propose a counterfactual analysis strategy supported by a causal path visualization and a counterfactual clustering visualization.}
Specifically, \pcheng{after experts intervene on a concept,} the causal path visualization helps reveal the induced causal effects, and the counterfactual clustering visualization helps shows which concepts or genes are affected.
The effectiveness of our proposed \pcheng{tool} is demonstrated through the two real-world case studies, the extraction of scientifically meaningful causal insights, and positive feedback from domain experts.
\pcheng{The source code is available at: https://github.com/hnu-vis/CELLens.}


In summary, the contributions of this work include:

\begin{itemize}[nosep]
\item\noindent{\textbf{A gene-similarity-aware causal graph visualization}  that preserves both the causal relationships between concepts and the similarities between genes.}
\item\noindent{\textbf{A counterfactual analysis strategy} that helps validate and refine causal graphs.}
\item\noindent{\textbf{Case studies} with real-world biological data to demonstrate the effectiveness of the proposed \pcheng{tool}.}
\end{itemize}

\section{Related Work}

This research focuses on utilizing interactive causal analysis to guide and refine virtual cells.
Accordingly, this section reviews the most pertinent literature across two primary domains: virtual cells and interactive causal analysis.

\subsection{Virtual Cells}
Virtual cell methods leverage \pcheng{machine learning models} to decode cellular mechanisms from \pcheng{gene expression.} 
While early black-box generative models (\eg, scVI~\cite{lopez2018deep}, scGen~\cite{lotfollahi2019scgen}, CellLM~\cite{zhao2023large}, and scGPT~\cite{cui2024scgpt}) achieve state-of-the-art predictive performance, their lack of biological interpretability hinders rigorous  mechanistic inference. 
To address this limitation, \pcheng{recent research focuses on two main paradigms:} interpretability-enhanced models and causal-driven models.

\textbf{Interpretability-enhanced models} \pcheng{improve interpretability by learning latent representations that separate cellular states according to distinct biological factors.}
\pcheng{Early efforts focused on separating basal states from intervention-induced states.}
\pcheng{CPA~\cite{lotfollahi2023predicting} addresses this separation by combining a basal representation with intervention changes to model intervention-induced states.
sVAE~\cite{lopez2023learning} further assumes that each intervention changes only a small subset of latent variables, while SAMS-VAE~\cite{bereket2023modelling} accounts for sample-specific variation in intervention-induced states.
Beyond these, biolord~\cite{piran2024disentanglement} separates cellular states by more attributes such as spatial, temporal, and disease states and
scDisInFact~\cite{zhang2024scdisinfact} further considers technical batch states.
However, these methods mainly focus on capture correlations, but not causal relationships for mechanistic analysis.}

\textbf{Causal-driven models} \pcheng{integrate predefined causal relationships into virtual cells to guide representation learning, response prediction, or counterfactual generation.}
\pcheng{GEARS~\cite{roohani2024predicting} and GeneCompass~\cite{yang2024genecompass} both use gene-gene relationships to model how interventions affect related genes.
CausCell~\cite{gao2025causal} further uses conditional diffusion models with causal relationships between biological concepts, enabling interpretable and controllable counterfactual generation of cellular states.} 
However, their performance is highly dependent on the quality of these predefined causal relationships. 
In practice, expert-provided causal graphs are often incomplete, whereas auto-mined graphs are often inaccurate, severely limiting their reliability and applicability.

\pcheng{In summary, these limitations suggest that causal relationships requires careful inspection before they are used to guide virtual cells.}
We therefore propose a human-in-the-loop visual analysis tool that supports expert-guided validation and refinement of causal relationships.

\subsection{Interactive Causal Analysis}
\pcheng{Interactive causal analysis incorporates domain experts into causal analysis workflows through visual interfaces, enabling them to explore, validate, and refine causal graphs.}
According to their analytical focus, existing works fall into two main categories: causal graph exploration and analysis, and causal relationship validation and refinement.

\textbf{Causal graph exploration and analysis} \pcheng{aims to help users interpret inferred causal graphs. 
Most existing tools present causal graphs as node-link diagrams, where nodes represent variables or concepts and directed edges encode causal links~\cite{forbes2017dynamic, yan2020silva, guo2023causalvis}.
Based on this, one group of work focuses on optimizing graph representations to reduce cognitive load, examining how visual encodings and layout algorithms affect user understanding~\cite{bae2017understanding, vo2020visual}.
Another group enriches causal graphs with additional analytical context.
For example, Xie~\etal~\cite{xie2020visual} use edge thickness to present causal-link uncertainty, Fan~\etal~\cite{fan2024visual} use comparative layouts to show differences across multiple outcome graphs, and Li~\etal~\cite{li2025causality} use a map-like metaphor to present hierarchical causal flows among topics.
However, these methods primarily support the interpretation of concept-level causal graphs.}
When applied to virtual cells, they are not sufficient to reveal the underlying concept-gene correspondences and gene-gene similarities.

\textbf{Causal relationship validation and refinement} \pcheng{focuses on evaluating and correcting inferred causal relationships. 
Although data-driven algorithms are computationally powerful, they can produce incorrect or missing causal links that conflict with prior knowledge.}
To support validation, interactive counterfactual analysis has been widely used to let users virtually intervene in variables and observe how other changes~\cite{kaul2021improving, borland2024using}.
Interactive systems such as Outcome-Explorer~\cite{hoque2021outcome} and VISPUR~\cite{teng2023vispur} leverage counterfactual reasoning to help analysts identify spurious associations and explain opaque algorithmic decisions.
Beyond counterfactual inspection, tools such as D-BIAS~\cite{ghai2022d} and Causalchat~\cite{zhang2025causalchat}, further integrate human-in-the-loop workflows, enabling domain experts to revise causal graphs by adding missing links or removing invalid connections.
\pcheng{However, when applied to virtual cells, current tools fail to sufficiently support joint validation the correctness of the high-level concept relationships and the underlying concept-gene correspondences.}

In summary, existing interactive causal analysis methods \pcheng{mainly focus on concept-level causal relationships, without incorporating the concept-gene correspondences and gene-gene similarities into causal exploration, validation, and refinement.}
To fill this gap, we propose \sys, a visual analysis system that combines gene-similarity-aware graph visualization with counterfactual analysis to support interactive causal relationship analysis for causal-driven virtual cells.

\addtocounter{figure}{0}
\begin{figure}[t]
    \centering
    \includegraphics[width=\linewidth]{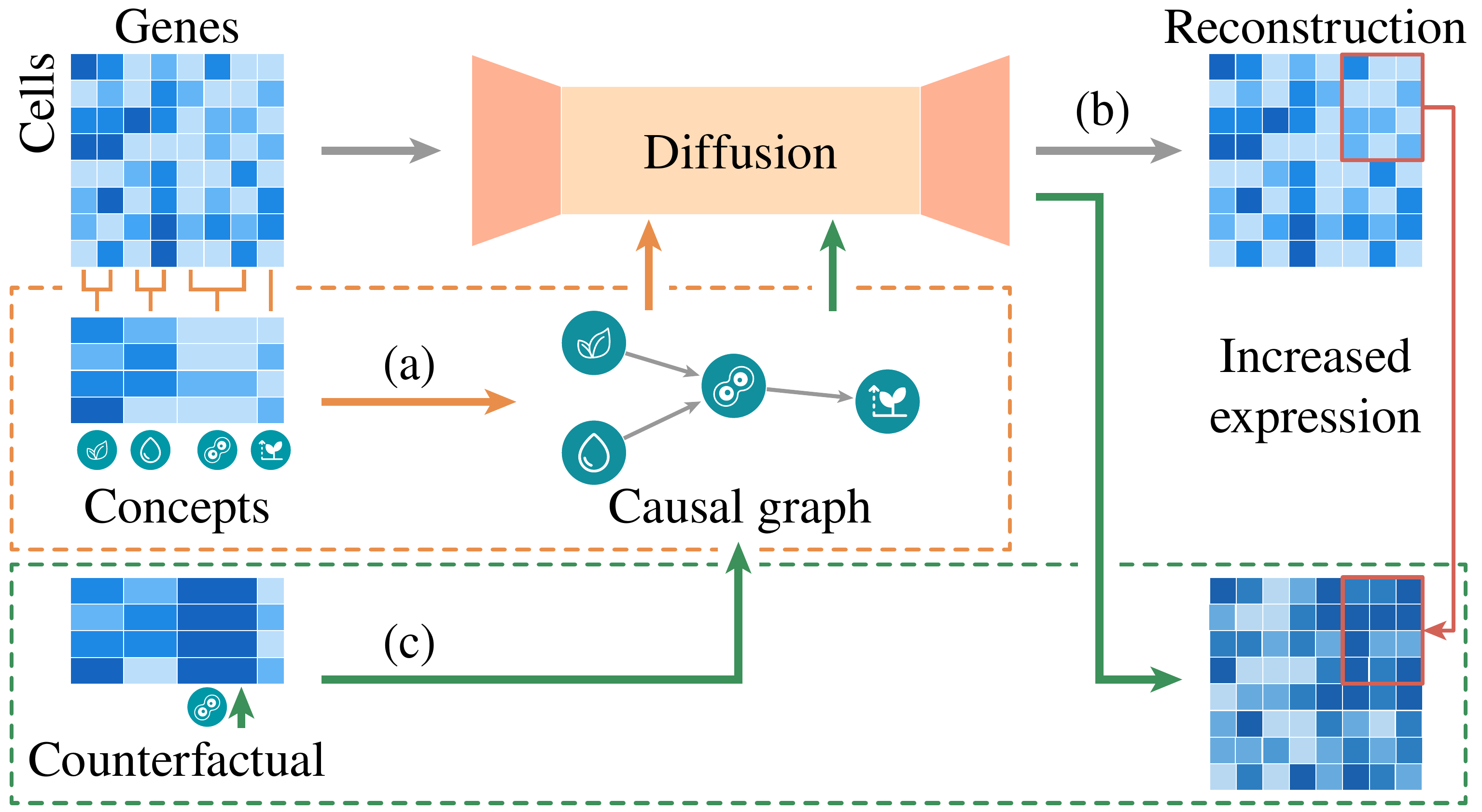}
    \caption{Causal-driven virtual cells training and analysis consists of three main steps: (a) causal graph construction, (b) virtual cell training, and (c) counterfactual generation.}
    \label{fig:background}
    \vspace{-2mm}
\end{figure}

\addtocounter{figure}{0}
\begin{figure*}[b]
    \centering
    \includegraphics[width=\linewidth]{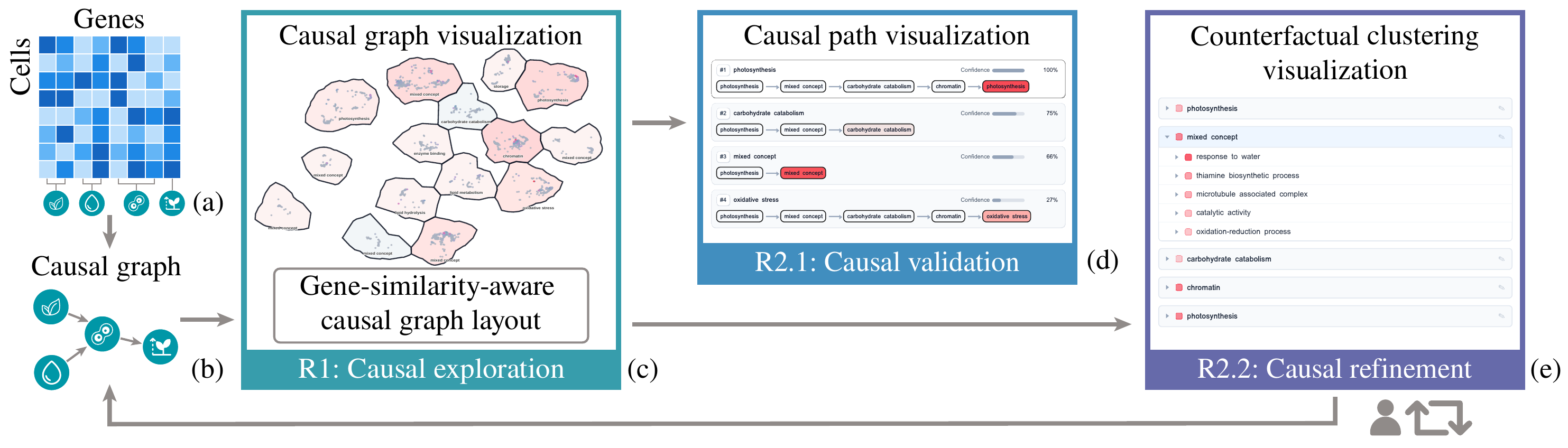}
    \caption{Method overview: (a-b) given a gene expression data, a group of concepts and their causal graph is constructed; (c)-(e) three visualizations to help explore the causal relationships and refine the causal relationships interactively.}
    \label{fig:pipeline}
\end{figure*}

\section{Background}
\label{sec:background}

Causal-driven virtual cells utilize gene expression data and underlying causal \pcheng{graphs} to simulate cellular behaviors and \pcheng{support mechanistic exploration.}
The computational pipeline of such models involves three key steps: causal graph construction, virtual cell training, and counterfactual generation.

\myparagraph{Causal graph construction.}
This step processes gene expression data to extract concepts and construct a concept-level causal graph for the virtual cell (Fig.~\ref{fig:background}(a)). 
The gene expression data is represented as a gene expression matrix, where rows correspond to cells and columns to genes. 
Each element in the matrix represents the expression level of a specific gene in a corresponding cell.
Following common practices~\cite{langfelder2008wgcna}, we use WGCNA to cluster genes into groups, treating each group as a high-level \textit{concept}.
Then, we construct a concept matrix that quantifies the cellular expression level corresponding to each concept.
To obtain the expression level for each concept, we extract rows of its member genes from the gene matrix to form a sub-matrix and utilize PCA to get the primary component as the expression level~\cite{langfelder2008wgcna}.
Based on this derived concept matrix, we apply the PC algorithm~\cite{spirtes2000causation} to infer the initial causal graph.
This algorithm first constructs a fully connected graph of variables and then systematically removes edges based on conditional independence tests. 

\myparagraph{Virtual cell training.}
\pcheng{After constructing the initial causal graph, we then subsequently train a causal-driven virtual cell.}
The virtual cell training follows a standard reconstruction paradigm: given the original gene expression matrix and the causal graph, the model reconstructs the corresponding gene expression matrix in an encoder-decoder manner.
In this work, we employ CausCell~\cite{gao2025causal}, a diffusion-based generative model, to train the virtual cell (Fig.~\ref{fig:background}(b)).
Rather than treating the extracted concept matrix as independent conditional inputs, CausCell utilizes a Structural Causal Model (SCM) to further integrate the causal graph.
SCM explicitly embeds the causal graph into the condition vectors, ensuring that the downstream generative process adheres to the concept-level causal relationships.

\myparagraph{Counterfactual generation.}
\pcheng{Counterfactual generation predicts how gene expression would change if a specific concept were intervened (Fig.~\ref{fig:background}(c)).}
To support this, continuous concept expression levels are first discretized into categorical states (\eg, low, \pcheng{medium,} high).
When simulating \pcheng{an intervention}, such as increasing the expression level of a specific concept, users alter its value in the concept matrix to construct an intervened one.
\pcheng{Guided by the embedded SCM, this localized intervention automatically propagates along the causal graph to update the expression of all downstream concepts.}
Conditioned on this intervened concept matrix, the generative model simulates the corresponding counterfactual gene expression matrix.
By comparing the generated gene expression matrix against the original gene expression matrix, experts can quantify the downstream response patterns of the intervention across affected genes.

\section{Requirement Analysis}
\label{sec:requirement}

This work was developed in close collaboration with three domain experts (B1--B3). 
B1 and B2 are biology researchers with over ten years of experience \pcheng{in interpreting gene regulatory relationships from gene expression data.}
\pcheng{Their analyses often involve causal graphs that represent candidate regulatory mechanisms mined from gene expression data.
However, these auto-mined graphs can be complex and inaccurate, making it difficult to understand and validate.}
B3 is a principal investigator specializing in causal inference for bioinformatics.  
B3 has long focused on improving causal discovery models, but still faces challenges caused by implausible causal relationships produced by auto-mining algorithms.
Therefore, they are exploring ways to better inject human biological expertise to improve the accuracy and completeness of these causal graphs. 
Over a 12-month collaboration, we conducted interviews every 2--4 weeks to collect requirements and gather feedback on our prototypes. 
\pcheng{Before this process, we obtained informed consent from all participants, and we did not collect any private or sensitive personal information from the experts.
According to our university’s IRB policy, this research is exempt from ethics review.}
Based on this iterative process, we summarize the following key analytical requirements.

\myparagraph{R1: Exploring the causal relationships at different levels}.
All experts stated that an initial step in virtual cell modeling is to comprehensively explore the auto-mined causal graphs. 
``Understanding how genes are grouped into concepts and how these concepts causally interact is crucial before we can trust any downstream simulation,'' B1 emphasized. 
\pcheng{Because the graph contains gene-gene similarities, gene-concept correspondences, and concept-concept causal relationships, experts require to explore them at different levels to help them assess whether the generated concepts are biologically meaningful and whether the inferred causal relationships are plausible.}

\myparagraph{R2: Validating and refining causal relationships}.
While data-driven algorithms are computationally powerful, the auto-mined causal graphs inevitably contain inaccuracies. 
``An algorithm might construct a causal link or group certain genes together in ways that violate established biological mechanisms,'' B3 noted. 
Therefore, domain experts need mechanisms to validate the generated concepts and causal relationships, and to refine them with biological expertise, thereby improving the biological reliability of virtual cells.

\textbf{R2.1:} \textit{Identifying biologically implausible causal relationships}.
Currently, validating auto-mined causal graphs is labor-intensive, \pcheng{as experts must repeatedly inspect gene-concept correspondences, causal links, and downstream responses across separate analysis steps.
Although counterfactual analysis tools can estimate the effects after intervening in a concept, their outputs are often not directly connected to the corresponding parts of the graph.
Therefore, experts need support for connecting counterfactual results back to the corresponding parts to check if the changes are biologically plausible.}
``I need to intervene in a specific concept and immediately observe which downstream concepts or genes are affected,'' B1 mentioned. 
This helps them determine whether the observed changes follow biologically plausible paths and identify implausible concept or causal relationships.

\textbf{R2.2:} \textit{Refining the causal relationships effectively and efficiently}.
After identifying implausible gene-concept correspondences or causal relationships, \pcheng{experts need to correct these errors to ensure the graph accurately reflects biological mechanisms.
Since such refinements can involve both concept-concept causal links and gene-concept correspondences, experts require effective and efficient mechanisms to directly update the auto-mined causal graph.}
``I need an intuitive way to remove an incorrect link, add a missing link, or correct an inappropriate gene grouping,'' B2 stated. 
Therefore, experts need support for directly updating the auto-mined causal graph according to their biological judgments, thereby reducing the effort required for refinement while improve both the accuracy and completeness of the causal graphs.

\section{{\sys} Visualization}
\label{sec:method}

Based on the identified requirements, we developed {\sys} to support interactive causal graph exploration and refinement for virtual cells.
Fig.~\ref{fig:pipeline} provides an overview of the developed method.
Given a collection of cells, their corresponding gene expression data (Fig.~\ref{fig:pipeline}(a)), and an initial causal graph (Fig.~\ref{fig:pipeline}(b)), these inputs are fed into the causal graph visualization to facilitate the exploration of causal relationships among concepts and similarities across genes (\textbf{R1}, Fig.~\ref{fig:pipeline}(c)).
Based on the exploration, experts perform virtual interventions on a specific concept.
\pcheng{They then use the causal path visualization to examine candidate causal paths associated with the intervention-induced responses (Fig.~\ref{fig:pipeline}(d)), and the counterfactual clustering visualization to inspect the internal composition of affected concepts, helping identify implausible causal relationships (\textbf{R2.1}, Fig.~\ref{fig:pipeline}(e)).}
Finally, experts explicitly adjust the concept-gene correspondences in the counterfactual clustering visualization and concept-concept causal \pcheng{relationships} in the causal visualization to inject their domain knowledge into the auto-mined causal graph (\textbf{R2.2}, Fig.~\ref{fig:pipeline}(e)).
\pcheng{Upon these visual modifications, the underlying model is dynamically updated to align with the corrected causal relationships.
This process iterates until the causal graph is sufficiently refined for virtual cell training.}

\subsection{Gene-similarity-aware Causal Graph Visualization}

To facilitate the causal graph exploration for virtual cells, it is essential to explore both the causal relationships between concepts and the similarities between genes simultaneously \pcheng{(\textbf{R1}).}
However, traditional causal graph visualizations typically rely on node-link diagrams that represent concepts as nodes~\cite{wang2015visual, xie2020visual, weinberg2025causality}.
Recent advanced causal graph visualizations utilize maps to better convey causal paths, such as CausalMap~\cite{li2025causality}.
However, these methods focus primarily on concept-level analysis and thus fail to preserve fine-grained gene-gene similarities.
According to the survey by Dennig~\etal~\cite{dennig2023fs},
one of the most effective \pcheng{ways} to reveal similarities is to employ projection methods (\eg, t-SNE~\cite{van2008tsne} and PCA~\cite{wold1987pca}) to create 2D scatterplots~\cite{PENG2025100234, TANG202448, chen2025human, chen2025interactive}.
Therefore, this motivates us to combine the causal graph visualizations with scatterplots to present both the causal relationships between concepts and the similarities between genes.

\pcheng{A straight way is to use a node-link diagram with circular nodes for concepts and place gene scatterplots inside each concept.
However, it will lead to overlap between neighboring projections when the edges are short, or waste screen space and reduce the effective area available for inspecting gene-gene similarities.}
Therefore, map-based causal graph \pcheng{visualization is} used due to its large spaces for each concept to locate the scatterplots \pcheng{while enabling the tracing of causal directions.}

\begin{figure*}[b]
    \centering
    \includegraphics[width=\linewidth]{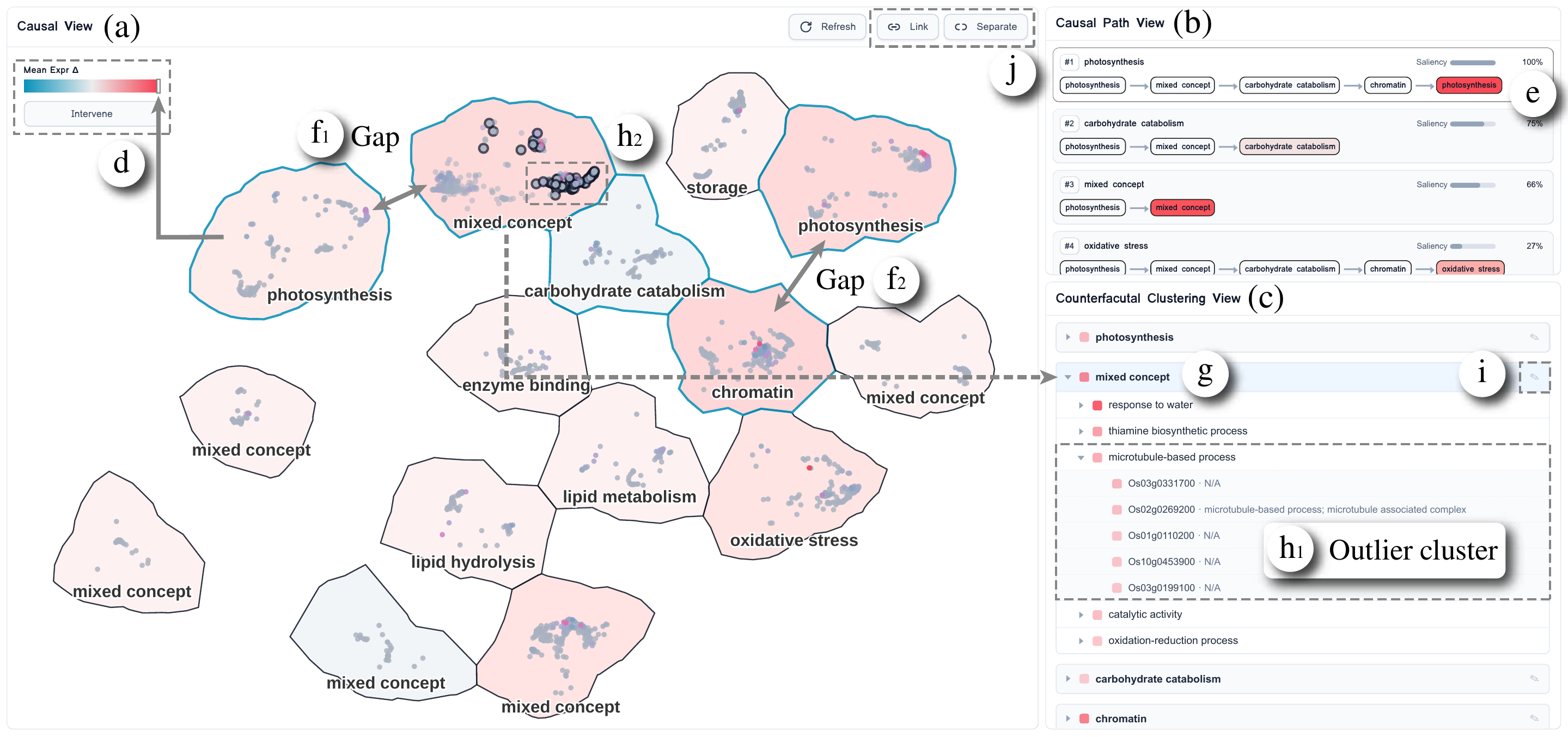}
    \caption{{\sys}: (a) a causal graph visualization facilitates the exploration of the concept-concept causal relationships, the gene-concept correspondences and the gene-gene similarities; (b) a causal path visualization helps validate the causal relationships; (c) a counterfactual clustering visualization helps explore and adjust the gene-concept correspondences thereby refining the causal relationships.}
    \label{fig:system}
\end{figure*}

In line with the above motivation, we developed a gene-similarity-aware causal graph visualization.
As shown in Fig.~\ref{fig:system}(a), biological concepts are visualized as polygonal regions.
Within each region, genes are embedded as a scatterplot to preserve fine-grained gene-gene similarities.
The causal relationships between these concepts are encoded via spatial border and directional ordering.
Specifically, bordering concept regions represent a causal link, with the left one acting as the source and the right one as the target.
To satisfy such encodings, it is required to: 1) preserve causal relationships between concepts along the horizontal direction, 2) preserve the similarities
between genes in each associated concept.
To achieve this, we propose a gene-similarity-aware causal graph layout algorithm.

\subsubsection{Gene-similarity-aware causal graph layout}

\myparagraph{Problem setting.}
Given a gene expression matrix $A \in \mathbb{R}^{n \times m}$, each column $a_i \in \mathbb{R}^n$ represents the expression values of the $i$-th gene across $n$ cells. 
Each column $a_i$ also serves as the feature of the $i$-th gene for concept grouping.
Specifically, $m$ genes are grouped into $p$ biological concepts $C = \{c_1, \dots, c_p\}$, governed by a directed causal graph $G = (C, E)$. 
Each edge $(c_i,c_j)\in E$ denotes a causal link from a source concept $c_i$ to a target concept $c_j$. 
Our goal is to compute a 2D projection $D \in \mathbb{R}^{m \times 2}$ that preserves both local gene-gene similarities and the global causal relationships. 
Each row $d_i=[d_i^{(x)}, d_i^{(y)}]$ of $D$ is the position of $i$-th gene in the 2D plane.
$D$ is expected to satisfy three constraints: 
1) maintaining similarities between genes within each polygonal region;
2) placing genes of the same concept within their associated polygonal region; 
and \pcheng{3) ensuring that each causal link is represented by two concept bordering regions, with the source concept placed to the left of its target concept.}
Moreover, to guarantee aesthetic and visual correctness, the layout is expected to be compact, and the polygonal regions of concepts should not overlap.
\pcheng{These jointly define an optimization problem, where gene positions, polygonal regions, and left-to-right causal direction need to be optimized simultaneously.}

\myparagraph{Optimization.}
\pcheng{Directly optimizing the above problem is highly nontrivial.}
This is primarily because \pcheng{the constraint of concept polygonal regions is associated with the convex hulls of their member genes, while} the computation of convex hulls is nested within the overall minimization process.
Any changes to the projected gene positions will alter the convex hulls of the corresponding gene concepts. 
Combining the layout minimization with dynamic convex hulls computation within a single-stage optimization problem leads to a substantial increase in computational complexity and numerical instability.
To address this issue, we propose a global-to-local optimization strategy that decouples the layout minimization from convex-hull-related computations. 
The global placement step determines the coarse spatial arrangement of concepts and their genes by using the genes of a concept to represent its convex hull.
The local placement step refines the positions of individual genes within each concept to ensure the convex-hull-related constraints.

\underline{\textit{Step 1: global placement}}.
In this step, we use the genes of a concept to represent its convex hull.
Specifically, we utilize the center of the genes of a concept to represent the center of its convex hull, and the hard constraints are converted into minimization objectives.
Moreover, motivated by the ability of t-SNE to preserve cluster separation, we compute the similarity based on the difference in similarity distributions between the high-dimensional and low-dimensional spaces. 
Accordingly, the problem can be formulated as the following constrained optimization:

\begin{equation}
\begin{aligned}
\label{eq:global-opt}
 \ell &=  KL(P || Q) + KL(P^c || Q^c) \\
 & \quad + \lambda_1 \sum_{(c_i, c_j) \in E} \max\left(0, \mu^{(x)}(c_i)-\mu^{(x)}(c_j)-\delta_x \right) \\
 & \quad + \lambda_2 \sum_{(c_i, c_j) \in E} \left\| \mu(c_i) - \mu(c_j) \right\|^2 \\
 & \quad + \lambda_3 \sum_{(c_i, c_j) \in S_{co}} -\left( \mu^{(y)}(c_i) - \mu^{(y)}(c_j) \right)^2
\end{aligned}
\end{equation}
The first two terms represent the preservation of gene-gene similarities and gene-concept correspondence, formulated as Kullback-Leibler (KL) divergence~\cite{meng2023class}.
\pcheng{Here, $P$ and $Q$ describe pairwise gene-gene similarities indexed by two genes $(i,j)$.}
Following t-SNE, the high-dimensional gene-gene similarity probability $p_{ij} \in P$ is computed using symmetric Gaussian distributions over the original features $a_i$~\cite{van2008tsne}. 
Its low-dimensional similarity probability $q_{ij} \in Q$ is modeled via a Student's t-distribution with one degree of freedom over the 2D projected position $d_i$:
\begin{equation}
    q_{ij} = \frac{(1 + \|d_i - d_j\|^2)^{-1}}{\sum_{k \neq l} (1 + \|d_k - d_l\|^2)^{-1}}.
\end{equation}

\pcheng{$P^c$ and $Q^c$ describe gene-concept correspondence indexed by a gene $i$ and a concept $k$.} 
The high-dimensional probability $p^c_{ik} \in P^c$ is defined as a fixed binary prior based on the initial clustering, formulated as an indicator function $p^c_{ik} = \mathbbm{1}(a_i \in c_k)$. 
To dynamically pull genes toward their corresponding concept centers in the 2D plane, the low-dimensional probability $q^c_{ik} \in Q^c$ is calculated using the Student's t-distribution between the projected gene $d_i$ and the concept center $\mu(c_k)$:
\begin{equation}
    q^c_{ik} = \frac{(1 + \|d_i - \mu(c_k)\|^2)^{-1}}{\sum_{j, l} (1 + \|d_j - \mu(c_l)\|^2)^{-1}}.
\end{equation}
Minimizing this divergence can update the spatial positions of both the genes and the concept centers, effectively aggregating member genes into cohesive visual clusters.

The remaining three terms impose geometric penalties on the concept centers
to ensure causal consistency and layout compactness:
1) The third term acts as a directional penalty to enforce a strict left-to-right spacing between causally connected concepts, where $\delta_x$ is a target horizontal margin.
$\mu(c_i)$ is the center of the genes of $i$-th concept.
$\mu^{(x)}(c_i)$ and $\mu^{(y)}(c_i)$ represent the position along the x-axis and y-axis, respectively.
2) The fourth term minimizes the spatial distance between causally connected concepts, encouraging linked concept regions to remain visually close and making the layout more compact.
3) The fifth term introduces a vertical repulsive penalty specifically for $S_{co}$, which denotes the set of concept pairs sharing a common target concept in the causal graph. 
This maximizes the vertical separation in $S_{co}$ to avoid overlapping. 
$\lambda_1$, \pcheng{$\lambda_2$}, and $\lambda_3$ are determined by a grid search to balance the magnitude difference among these terms.
Similar to t-SNE, Eq.~(\ref{eq:global-opt}) is optimized with gradient descent.
\pcheng{The optimization details are provided in Appendix A.}

\underline{\textit{Step 2: local placement}}.
While the first step successfully positions the concepts in a causal layout, the non-overlapping constraint between convex hulls may be violated.
In such cases, genes near the boundaries of adjacent concepts may still overlap.
To address this, this step applies a force-directed fine-tuning to the position of genes in the 2D plane to separate boundary genes without disrupting the global causal topology established in the first step.

To separate these overlapping genes, we introduce a center-attractive loss. For any two genes $a_i \in c_k$ and $a_j \in c_l$ belonging to different concepts ($c_k \neq c_l$), if their Euclidean distance is less than a margin $\epsilon$, they are penalized to pull back toward their respective concept centers, $\mu(c_k)$ and $\mu(c_l)$. The formulation is as follows:
\begin{equation}
\begin{aligned}
\ell_{\mathrm{pull}} = \sum_{c_k \neq c_l} \sum_{\substack{a_i \in c_k,a_j \in c_l}}  & \mathbbm{1} \left(\| d_i - d_j \| \le \epsilon \right) \\
&  \quad \cdot \left( \| d_i - \mu(c_k) \|^2 + \| d_j - \mu(c_l) \|^2 \right).
\end{aligned}
\end{equation}
Minimizing $\ell_{\mathrm{pull}}$ ensures that boundary genes violating the margin $\epsilon$ are dynamically drawn toward their own concept's center. 
Consequently, this two-step optimization produces distinct, non-overlapping \pcheng{convex hulls,} providing a clear and mathematically rigorous layout for subsequent visual analysis. 

Once the optimization is complete, we calculate the \pcheng{polygonal regions} of each concept based on the GMap algorithm~\cite{gansner2010gmap}. 
Standard GMap computes a Voronoi diagram and introduces invisible random points around the given points to generate finite outer borders~\cite{aurenhammer1991voronoi}. 
Voronoi cells sharing the same attribute (\ie, concept) are then merged to form contiguous polygonal regions. 
However, standard GMap forces all polygonal regions to be bordering, failing to well represent the causal relationships in causal graphs. 

To accurately represent the causal graphs, we introduce a margin-injection method.
Initially, we apply the standard Voronoi partitioning to generate a continuous map and identify the shared boundaries between all spatially bordering regions. 
Subsequently, for concept polygonal regions without direct causal links, we \pcheng{add} invisible virtual points along their shared boundaries. 
By recomputing the Voronoi diagram with these virtual points, the algorithm generates spatial margins that explicitly separate the causally independent concepts.
\pcheng{The details of region construction are provided in Appendix B.}

Furthermore, this margin-injection method inherently supports interactive causal editing. 
By dynamically inserting or removing these points, the underlying Voronoi partitioning updates efficiently, providing real-time visual feedback when experts link or separate concepts.

\subsubsection{Interaction}
We provide several interactions to help better explore and refine the causal graphs in the causal graph visualization.

\myparagraph{Semantic annotation}.
\pcheng{It is crucial for users to intuitively understand the biological semantics of the concepts (\textbf{R1}).
To achieve this, for each concept, we input all its member genes into Gene Ontology (GO) enrichment analysis,} a standard technique for deriving representative biological semantics from a group of genes~\cite{ashburner2000gene}.
\pcheng{This analysis maps these genes against the GO database~\cite{sakai2013rice, kawahara2013improvement}, identifying semantic terms (\eg, ``growth'' or ``metabolism'') that appear more frequently than expected by chance.}
However, because these raw terms are often redundant and highly granular, \pcheng{they are not intuitive for visual exploration.}
\pcheng{Therefore, we use an LLM to summarize these GO terms into single concise concept names as initial semantic summaries.
Users further examine whether each name is biologically meaningful and revise it when necessary.
The specific LLM and prompt used in our implementation are provided in Appendix C.
}

\myparagraph{Causal editing.}
\pcheng{It is essential for users to directly refine concept-concept causal relationships in the \emph{Causal Graph Visualization} (\textbf{R2.2}). 
To achieve this, we support region-based causal editing. 
By clicking and dragging a concept region, users can adjust its horizontal position to revise the direction of related causal links, or reposition regions to fine-tune the global layout.}
Additionally, users can \pcheng{add} or \pcheng{remove} causal links by multi-selecting two regions and clicking the ``link'' or ``separate'' buttons (Fig.~\ref{fig:system}(j)). 
\pcheng{These direct interactions update the underlying causal graph, allowing domain experts to correct biologically implausible concept-concept causal relationships.}

\myparagraph{\pcheng{Counterfactual generation}}.
\pcheng{To validate causal relationships, it is crucial for experts to understand how a change in one concept affects the others (\textbf{R2.1}).
Counterfactual generation allows users to initialize an intervention on a concept as the starting point for this process.}
Users first select a concept region and adjust the legend slider to specify a target expression change, \pcheng{where moving the slider to the right increases expression and moving it to the left decreases it (Fig.~\ref{fig:system}(d)).}
Upon clicking the ``intervene'' button, the system generates the counterfactual result and quantifies the response of each gene $a_i$ as its expression change $\Delta(a_i)$.
\pcheng{Subsequently, to capture the dominant expression response of each concept, we compute the aggregate change $\Delta(c_i)$ of each concept by averaging the expression changes of its top 5\% member genes ranked by absolute expression change~\cite{subramanian2005gene}.}
The fill color of each concept region reflects this aggregate change, \pcheng{with red indicating an increase, blue a decrease, and intensity showing the magnitude.}
Meanwhile, individual gene points use the same color scale, \pcheng{providing gene-level responses for downstream visualizations.}

\subsection{Causal Path Visualization}
\pcheng{We provide a way to identify biologically implausible causal relationships (\textbf{R2.1}) by allowing users to examine paths at the concept level.
A causal path is defined as a sequence of concepts connected by causal links, indicating a possible route from the intervened concept to an affected concept.}
In this section, we describe how these paths are extracted and visually encoded to facilitate expert validation.

\myparagraph{Causal path extraction.}
Based on the counterfactual generation, \pcheng{we extract candidate causal paths from the intervened concept $c_s$ to all affected concepts $c_t$ in the causal graph and prioritize them.}
\pcheng{To avoid missing potentially relevant paths, we traverse an augmented causal graph $G'=(C, E')$ that includes both constructed directed causal links and statistically associated links.
These statistically associated links are undirected edges inferred from the data using the PC algorithm.}
For each affected concept $c_t$, we employ breadth-first search to extract the shortest path from $c_s$, with its length denoted as $h$.
Subsequently, these extracted paths are ranked using a \pcheng{saliency} score formulation:
\pcheng{\begin{equation}
\mathrm{Saliency}(c_t) = \frac{|\Delta(c_t)|}{h}.
\end{equation}}
Dividing by $h$ penalizes longer paths, reflecting the biological principle that signals naturally weaken as they pass through multiple steps~\cite{cowen2017network}.
This approach helps highlight the most \pcheng{plausible} causal paths while filtering out \pcheng{implausible} ones.

\myparagraph{Visual encoding.}
Based on the calculated scores, the \emph{Causal Path Visualization} \pcheng{presents the candidate paths in descending order of saliency (Fig.~\ref{fig:system}(b)).}
\pcheng{The view presents these paths as a sorted list of cards. 
Inside each card, the path is displayed as a left-to-right sequence of concepts. 
The last concept in each path represents the concept affected by the intervention and is therefore color-coded to indicate its expression change $\Delta(c_t)$, following the same color scale as the \emph{Causal Graph Visualization}.
A bar on the right side visualizes the path's saliency score. 
When users click a path card, the corresponding causal path is highlighted in the \emph{Causal Graph Visualization} by changing the border colors of the involved concept regions.
By examining these cards, experts can they can filter out implausible causal paths and discover potential paths for further validation and refinement.}

\subsection{Counterfactual Clustering Visualization}
\pcheng{In addition to causal path validation, users need to inspect gene-level responses within each concept (\textbf{R2.1}) and refine concept-gene correspondences (\textbf{R2.2}).}
This section describes the counterfactual-aware clustering approach used to support these inspections and refinements.

\myparagraph{Counterfactual-aware clustering.}
Since a single concept often contains numerous genes, individual inspection is cognitively demanding. 
To facilitate analysis, we perform counterfactual-aware clustering within each concept \pcheng{to organize genes into interpretable groups.}
\pcheng{This clustering should capture how genes respond to the intervention while preserving the biological meanings needed for interpretation.
In particular, we characterize each gene $a_i$ using both its counterfactual response $\Delta(a_i)$ and its semantic representation $g_i$.
Specifically, $g_i$ is constructed in two steps: we first collect all unique GO terms within the concept to form a unified semantic feature space, and then encode the GO-derived representation of each gene in this space as a multi-hot binary vector.}
The combined feature vector is then formulated by concatenating the semantic representation and the counterfactual response:
\begin{equation}
f_i = \bigl[ g_i,\; \Delta(a_i) \bigr].
\end{equation}
\pcheng{To balance the numerical scales of the two distinct features,} we subsequently L2-normalize this vector to $\tilde{f}_i = f_i / \|f_i\|_2$. 
Within each concept, genes are then grouped using the K-means algorithm~\cite{macqueen1967some}, \pcheng{where $K$ is automatically selected as the elbow point of the within-cluster sum of squares curve~\cite{satopaa2011finding}.}
\pcheng{This dual-feature representation encourages clustered genes to share similar response patterns and biological semantics. 
Finally, we name each cluster using the same GO enrichment and LLM summarization as in the \emph{Causal Graph Visualization}.}

\begin{figure}[t]
    \centering
    \includegraphics[width=\linewidth]{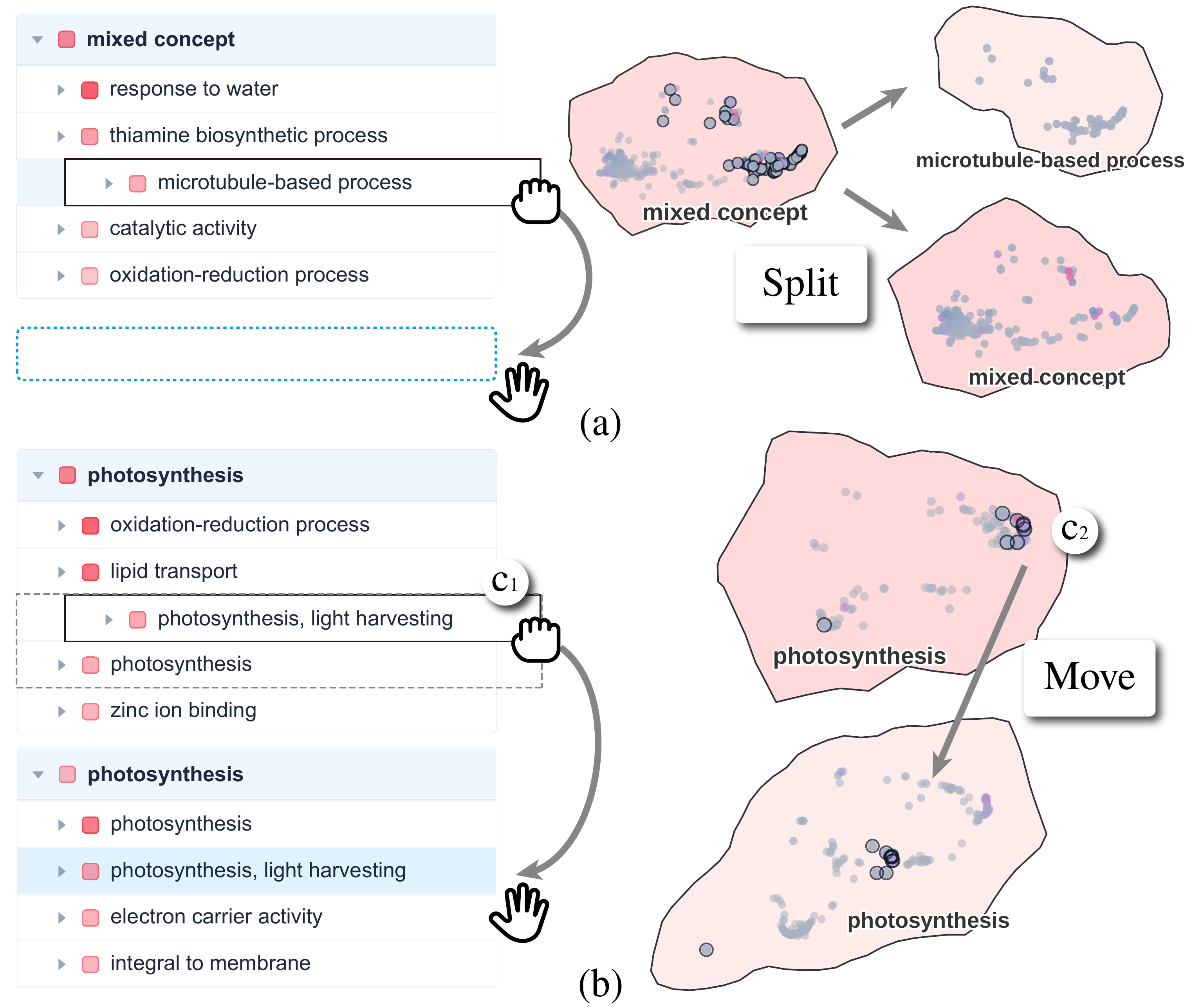}
    \caption{Drag and drop for adjusting the gene-concept correspondences.}
    \label{fig:case1-step2}
\end{figure}

\myparagraph{Visual encoding.}
Based on the clustering results, the \emph{Counterfactual Clustering Visualization} employs a nested ``concept--cluster--gene'' visual structure (Fig.~\ref{fig:system}(c)). 
Each row begins with a colored square indicating the expression change \pcheng{after intervention.}
\pcheng{For concepts and clusters, the square encodes the aggregate change of their member genes, computed in the same way as in the \emph{Causal Graph Visualization}; for individual genes, it encodes the gene-level change.}
Rows are labeled with summarized names for concepts and clusters, while individual genes are identified by their global IDs~\cite{maglott2005entrez}. 
To maintain a clear display, each cluster shows only its top five representative genes. 
This group-based design allows users to \pcheng{evaluate semantic consistency at the cluster level} rather than examining individual genes.

\pcheng{If the response pattern or the summarized name of a cluster conflicts with its parent concept, users can adjust the assignment via drag-and-drop.} 
Specifically, dragging a cluster into empty space extracts it to form a new concept (Fig.~\ref{fig:case1-step2}(a)). 
Alternatively, dropping the cluster into another concept reassigns its membership (Fig.~\ref{fig:case1-step2}(b)). 
Once the refined concept structure is considered appropriate, users can click the edit button to rename the concept (Fig.~\ref{fig:system}(i)). 
\pcheng{This interactive process refines the causal relationship for virtual cell retraining.}

\section{Evaluation}
\label{evaluation}

To demonstrate the effectiveness of {\sys} for facilitating the exploration, validation, and refinement of causal relationships in virtual cells, we performed two case studies focusing on the refined virtual cell and the identified key regulatory genes.

\subsection{Case Study}
\pcheng{The case study was conducted with B1 involved in the requirements analysis. }
B1 aimed to uncover the underlying mechanisms of rice.
Therefore, B1 used {\sys} to analyze rice gene expression data.
In the case study, to allow B1 to focus more on analysis, we used the pair analytics protocol, in which we handled the tool's navigation~\cite {arias2011pair, chen2025human}.
Before this process, we obtained informed consent from B1, and we did not collect any private or sensitive personal information from B1.

\subsubsection{\pcheng{Virtual Cell Refinement}}

\myparagraph{Preliminary}.
To uncover the mechanisms of rice, B1 aimed to train a causal-driven virtual cell specific to this species. 
He began by curating a recent rice single-cell dataset~\cite{wang2025single}, which contains 116,564 cells from eight distinct organs.
Since B1 focused on gene functions, he extracted only the gene expression data to construct the model. 
To mitigate the noise from lowly expressed genes, B1 followed a standard data preprocessing pipeline~\cite{gao2025causal}, retaining 115,393 high-quality cells and the top 2,000 highly expressed genes. 
\pcheng{Based on the progress detailed in Sec.~\ref{sec:background}, these genes were grouped into 15 high-level concepts, serving as the foundational nodes for the initial concept-level causal graph.}

\begin{figure}[t]
    \centering
    \includegraphics[width=\linewidth]{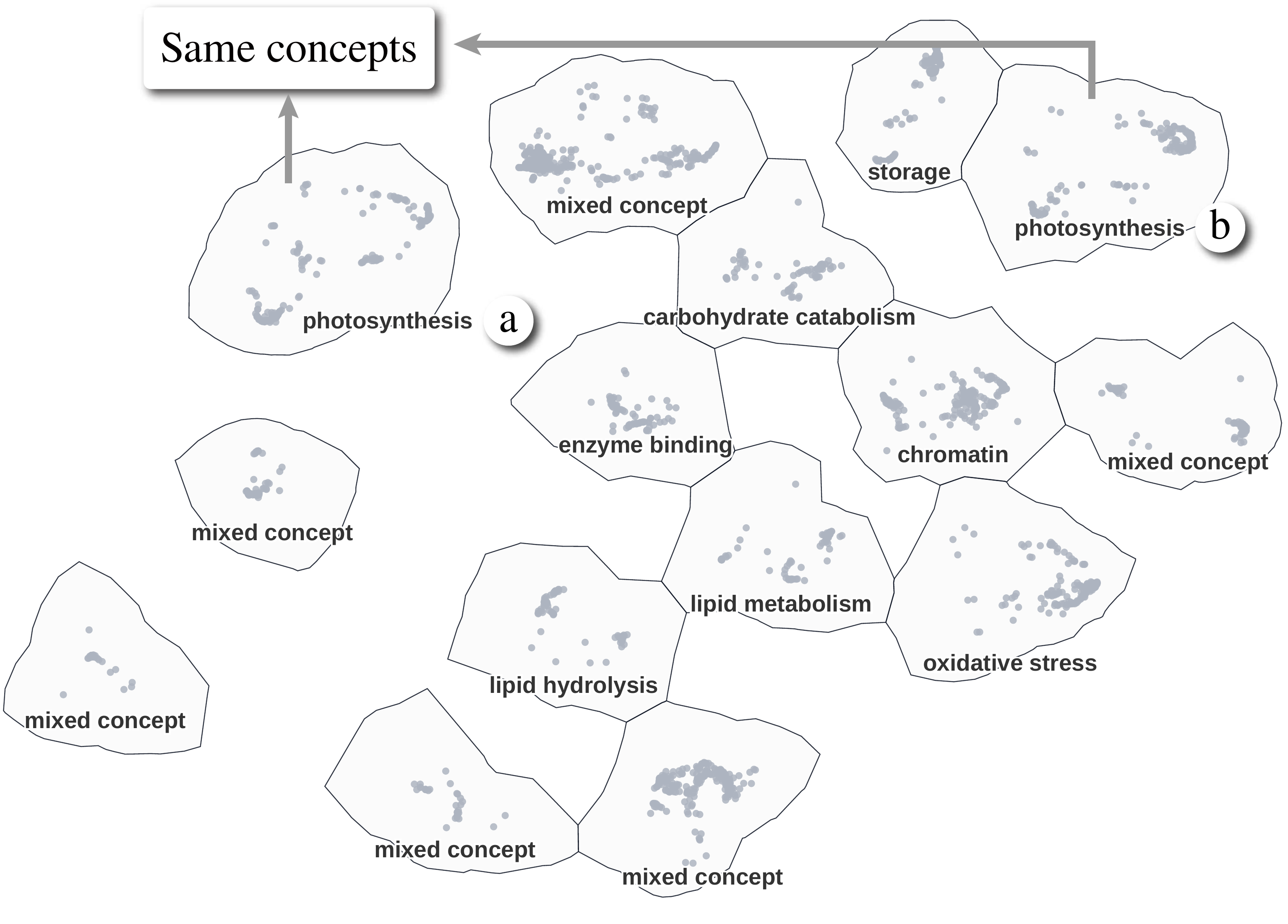}
    \caption{The initial causal graph visualization.}
    \label{fig:case1-overview}
\end{figure}

\myparagraph{Overview}.
B1 began his analysis with the causal graph visualization. 
Initially, the concept regions were colorless, indicating that no counterfactual interventions had been applied (Fig.~\ref{fig:case1-overview}). 
He noted that most concepts in the initial causal graph were connected, \pcheng{while several isolated concepts were located at the periphery.} 
\pcheng{Wondering why these isolated concepts were not connected to the others, B1 decided to check them first.}
Among them, he observed two distinct concepts with the same name, ``\textit{photosynthesis}'' \pcheng{(Figs.~\ref{fig:case1-overview}(a) and~\ref{fig:case1-overview}(b)).} 
\pcheng{The duplicated naming prompted B1 to examine whether they represented the same or different biological semantics.}
He first focused on \textit{photosynthesis} A \pcheng{(Fig.~\ref{fig:case1-overview}(a))}, which lacked any direct connections to the rest of the graph.
\pcheng{To understand how the other concepts were affected by \textit{photosynthesis} A,
B1 applied a counterfactual intervention to increase its expression level (Fig.~\ref{fig:system}(d)).
The virtual cell then generated concept-level changes across the graph, revealing the responses of other concepts.}

\begin{figure*}[b]
    \centering
    \includegraphics[width=\linewidth]{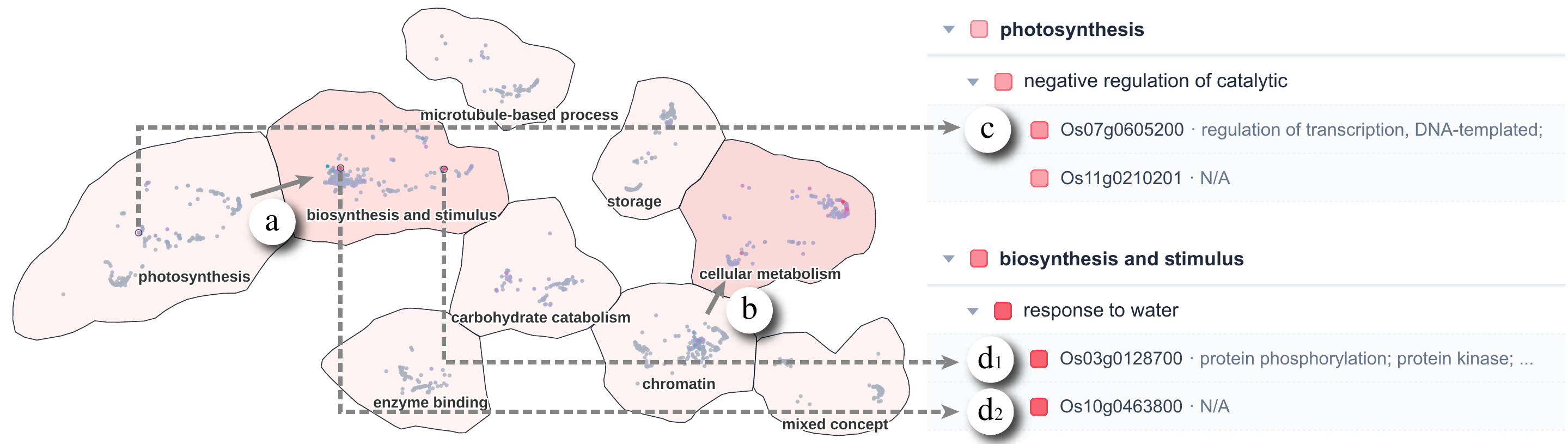}
    \caption{Refining the causal relationships and updating the counterfactual generation result.}
    \label{fig:case2}
\end{figure*}

\myparagraph{Disentangling mixed concepts}.
Upon update, the color of concept regions changed, \pcheng{reflecting the magnitude of their changes} in response to the intervention (Fig.~\ref{fig:system}).
\pcheng{In the causal path visualization, the top-ranked causal path with a high saliency score originated from \textit{photosynthesis} A caught B1's attention.
This causal path passed through several intermediate concepts, and pointed to \textit{photosynthesis} B (Fig.~\ref{fig:system}(e)).
To examine how the concepts in this path changed after the intervention, B1 checked the color of their corresponding regions.
He noted that most regions were shown in darker red, indicating relatively large positive changes, which suggested that the concepts along this path were strongly associated with the intervention response.
However, this path was misaligned with the causal graph via two missing links (Figs.~\ref{fig:system}(f$_1$) and (f$_2$)).
B1 therefore decided to examine these two missing links first to understand why they were missing in the causal graph.}

Investigating the first missing link (\textit{photosynthesis} A $\rightarrow$ mixed concept) (Fig.~\ref{fig:system}(f$_1$)), \pcheng{B1 found that the target concept was named ``mixed''.}
\pcheng{Recognizing that such semantic ambiguity could affect the causal discovery algorithm~\cite{scholkopf2021toward},} he hypothesized that the missing link might be caused by the mixed semantics of the target concept.
\pcheng{To determine what components contributed to the mixed semantics,} B1 analyzed this concept's internal clusters.
He found that most clusters' names were associated with biosynthesis and stimulus activities (\eg, response to water, thiamine biosynthetic process, and catalytic activity) (Fig.~\ref{fig:system}(g)). 
However, he noticed an outlier cluster named ``microtubule-based process,'' \pcheng{which is related to cytoskeleton organization} (Fig.~\ref{fig:system}(h$_1$)).
B1 explained that this \pcheng{cytoskeleton-related cluster was semantically distinct from the biosynthesis- and stimulus-related clusters, making it unreasonable to cluster them together.}

\pcheng{To further examine whether the genes in this outlier cluster differed from those in other clusters, B1 selected this cluster to inspect the projection of its member genes.}
He observed that its genes were spatially segregated on the right side of the concept region, indicating their difference to other clusters (Fig.~\ref{fig:system}(h$_2$)). 
\pcheng{Additionally, B1 expanded this outlier cluster to check the annotations of its member genes.}
He noted that most genes within this cluster were annotated as N/A (Fig.~\ref{fig:system}(h$_1$)), suggesting that limited annotation information may have contributed to its misclassification. 
\pcheng{B1 therefore decided to separate the ``microtubule-based process'' cluster from the mixed concept. 
After checking the names of other concepts, he found no existing concept that could appropriately accommodate this cluster.}
He therefore dragged it into the empty space in the counterfactual clustering visualization (Fig.~\ref{fig:case1-step2}(a)), extracting it into an independent concept. 
\pcheng{After separating this outlier cluster, the remaining concept showed clearer semantics related to biosynthesis and stimulus activities.}
Therefore, he then renamed the remaining concept to ``\textit{biosynthesis and stimulus}'' and explicitly added the missing causal link from \textit{photosynthesis} A (Fig.~\ref{fig:case2}(a)).

\myparagraph{\pcheng{Correcting a misidentified downstream concept}}.
B1 then investigated the second missing link (\textit{chromatin} $\rightarrow$ \textit{photosynthesis} B) (Fig.~\ref{fig:system}(f$_2$)). 
From a biological perspective, he considered that photosynthesis is unlikely to be a direct downstream concept of chromatin~\cite{jan2022retrograde}.
\pcheng{Given the noticeable change in \textit{photosynthesis} B after the intervention, B1 suspected that the concept might still contain components responsive to chromatin, while its name might not accurately reflect its internal composition.}
He therefore examined the internal clusters of \textit{photosynthesis} B to inspect its semantic composition.
He found that there were indeed two clusters related to photosynthesis (Fig.~\ref{fig:case1-step2}(c$_1$)).
\pcheng{However, their colors were light, indicating relatively small changes after the intervention.}
\pcheng{For further analysis, B1 checked the gene projection of these two clusters.}
He observed that although their projections were concentrated, they contained only a small number of genes (Fig.~\ref{fig:case1-step2}(c$_2$)), \pcheng{suggesting that they were unlikely to represent the primary semantics of the concept.}
\pcheng{Considering that GO-enrichment-based naming mainly relies on frequent annotation patterns, B1 inferred that these small but semantically consistent clusters likely biased the concept name, contributing to its misidentification.}

To correct this \pcheng{semantic misassignment}, he dragged the two photosynthesis clusters into \textit{photosynthesis} A at the start of the path, merging them with the more biologically coherent upstream photosynthesis concept (Fig.~\ref{fig:case1-step2}(b)). 
After this reassignment, B1 checked the remaining clusters and found that they mainly represented cellular metabolism-related activities \pcheng{(\eg, oxidation-reduction process, lipid transport, and zinc ion binding).}
Consequently, he renamed the refined concept to ``\textit{cellular metabolism}'' and explicitly added the missing causal link from \textit{chromatin} to it (Fig.~\ref{fig:case2}(b)).

\myparagraph{Verifying the refined model}.
\pcheng{After completing the causal relationship refinements, B1 retrained the virtual cell using the updated causal relationships.
In the previous exploration, he inferred that \textit{photosynthesis} should positively regulate \textit{biosynthesis and stimulus}, as the downstream concept region turned red after the intervention.
To examine whether the updated causal relationships helped the model learn this expected positive influence, he conducted a counterfactual experiment on the test set by comparing three conditions: the control condition without intervention, the initial model after increasing \textit{photosynthesis}, and the refined model after the same intervention. 
He then compared the mean gene expression of \textit{biosynthesis and stimulus} across these conditions.}

\pcheng{Generally, the refined model achieved the highest mean expression of 0.162 (95\% CI [0.158, 0.165]), the control condition got the second at 0.156 (95\% CI [0.151, 0.160]) and the initial model got the lowest at 0.131 (95\% CI [0.127, 0.134]).
The Friedman test~\cite{friedman1937use} indicated that the difference among the three conditions was statistically significant ($p<0.001$).
This result matched B1's expectation, indicating that the refined model better reflected the hypothesized positive regulatory relationship between \textit{photosynthesis} and \textit{biosynthesis and stimulus}.}

\subsubsection{Key Regulatory Gene Identification}

\myparagraph{Identifying regulatory genes along the pathway}.
After refining the causal graph and retraining the virtual cell, \pcheng{B1 aimed to identify candidate regulatory genes associated with the refined causal path from \textit{photosynthesis} to \textit{cellular metabolism}, which passed through several intermediate concepts.} 
\pcheng{Assuming that genes involved in regulating this path would show stronger responses to intervention, he performed a counterfactual intervention on the start concept \textit{photosynthesis}.}
In the updated scatterplot of the causal graph, some gene points appeared darker than others, indicating larger expression changes.
B1 noted that these darker genes were mainly located in concepts along the path, specifically within the \textit{biosynthesis and stimulus} and \textit{cellular metabolism} concepts (Fig.~\ref{fig:case2}).

\pcheng{B1 initially aimed to identify a regulatory gene at the start of the causal path.
Therefore, he examined candidate regulatory genes within the source concept, \textit{photosynthesis}.
Within this region, he observed several darker gene points. 
To further inspect these genes, he expanded the darkest red cluster in \textit{photosynthesis} and identified the gene with the darkest red color,  Os07g0605200 (Fig.~\ref{fig:case2}(c)).
Noting that it is annotated with transcription regulation and DNA-templated processes, which are relevant to photosynthesis-related regulation~\cite{burgess2017transcriptional}, B1 hypothesized that this gene might act as an upstream regulator involved in the downstream changes along this causal path.}
He then examined the downstream concept, \textit{biosynthesis and stimulus}, identifying two genes shown in clearly darker red: Os03g0128700 (Fig.~\ref{fig:case2}(d$_1$)) and Os10g0463800 (Fig.~\ref{fig:case2}(d$_2$)). 
\pcheng{Based on these visual and annotation cues, B1 inferred that Os07g0605200 could be an upstream regulatory gene associated with the link from \textit{photosynthesis} to \textit{biosynthesis and stimulus}, potentially influencing the two downstream responsive genes.}

\myparagraph{Regulatory gene verification}.
\pcheng{To check these findings with existing biological evidence, B1 consulted recent literature.
A recent study reported a regulatory relationship between Os07g0605200 and Os10g0463800 in the control of grain chalkiness~\cite{wang2026osmads18}, which provided gene-level support for B1's inference.
B1 noted that this gene-level evidence was consistent with the inferred concept-concept relationship and concept-gene correspondences. 
Specifically, Os10g0463800 is related to grain chalkiness, which has been associated with starch biosynthesis under environmental stimuli~\cite{yamakawa2007comprehensive}.
This supported the correspondence between Os10g0463800 and the \textit{biosynthesis and stimulus} concept.
Moreover, starch biosynthesis relies on carbon and energy supplied by upstream photosynthesis~\cite{zeeman2010starch}, providing support for the inferred positive relationship from \textit{photosynthesis} to \textit{biosynthesis and stimulus}.}

\pcheng{In contrast, B1 found no prior reports linking Os07g0605200 with Os03g0128700.
Since regulatory responses can vary across biological conditions, he considered Os03g0128700 a potentially novel downstream gene.}
Following this exploratory rationale, he further inspected highly responsive genes within the \textit{cellular metabolism} concept.
Ultimately, B1 selected these newly identified genes as candidates for future experimental analysis.

\section{Expert Feedback and Discussion}
\label{sec:discuss}

After the case study sessions, we conducted \pcheng{semi-structured interviews with five domain experts (B1--B5)} to gather feedback.
\pcheng{Among them, B1--B3 had participated in the requirement analysis and iterative tool development, while B4--B5 were newly invited domain experts who had not been involved in the previous process.}
Since \pcheng{B2--B5} were not involved in the case study, we began by providing them with a 30-minute introduction to the system and the case study.
Following the introduction, each subsequent interview lasted between 30 and 50 minutes.
\pcheng{The interviews were guided by several questions: whether the causal graph visualization helped experts understand concept-concept causal relationships, gene-concept correspondences and gene-gene similarities; whether the causal path and counterfactual clustering visualizations helped them validate causal relationships and locate relationships requiring refinement; and whether the refinement interactions matched their expected causal refinement process.}
\pcheng{Beyond these questions, we also invited experts to share additional benefits and useful suggestions for improvement.}
\pcheng{Before this process, we obtained informed consent from all participants, and we did not collect any private or sensitive personal information from the experts.}
Overall, the experts provided positive feedback regarding the \pcheng{effectiveness} of {\sys} in \pcheng{exploring, validating, and refining the causal relationships in virtual cells.}
Based on their feedback and our experience developing the tool, we also identified several limitations that require further investigation.

\subsection{\pcheng{Effectiveness}}

\myparagraph{Transparent exploration of cellular mechanisms}.
All experts appreciated the gene-similarity-aware causal graph, \pcheng{emphasizing that it successfully preserves both concept-concept causal relationships, gene-concept correspondences and gene-gene similarities.} 
\pcheng{This design supports mechanism exploration from both the concept and gene levels.}
B2 noted, ``This design allows us to easily trace how an intervention in one concept propagates along causal paths to affect the downstream concepts.''  \pcheng{Regarding gene-gene similarities within each concept,} B2 specifically praised the spatial arrangement within each concept \pcheng{region}: ``By looking \pcheng{at} the distances between genes inside a concept, I can easily understand its semantic purity.''
\pcheng{B4 further emphasized that linking concept and gene-level evidence is important, because a causal relationship alone is often insufficient for biological interpretation.}
\pcheng{This transparent layout helps biologists effectively identify biologically implausible concept formations before executing complex simulations.}

\myparagraph{Interactive refinement of causal relationships}.
B3 particularly praised how the tool integrates counterfactual \pcheng{generation} with interactive visual editing.
\pcheng{The experts pointed out that errors in virtual cells may arise from incorrect or missing concept-concept causal relationships and inaccurate concept-gene correspondences, and {\sys} provides direct interactions to refine both types of structures.}
\pcheng{For concept-concept causal relationships, the experts valued the immediate visual feedback after the intervention action, as updated counterfactual colors and causal paths help them judge whether the modified structure produces a more plausible downstream response.}
\pcheng{For concept-gene correspondences,} B3 noted, ``In traditional tools, fixing an error requires manual adjustment, which is extremely slow. In {\sys}, the intuitive drag-and-drop interaction allows me to easily reassign misclassified clusters to restore correct concept-gene correspondences or extract them to form new concepts.''
\pcheng{B5 further noted that this human-in-the-loop process allows experts to use their domain knowledge to find implausible relationships, check the results through counterfactual feedback, and edit them directly.}

\myparagraph{Generalization to other \pcheng{domains}}.
While the case study focused on rice cell mechanisms, B2 emphasized that the core framework, specifically the causal graph visualization and counterfactual generation, can be applied to biological problems beyond rice cell analysis.
``The fundamental definition of these causal concepts is universal,'' B2 elaborated. ``This system can be \pcheng{adapted} to explore other complex regulatory networks, such as cell-cell communication in tumor microenvironments or the driving factors behind immune cell exhaustion.''
\pcheng{B4 further noted that the proposed tool can also be utilized in other domains with similar causal analysis needs.
This is because the analysis in {\sys} is built around high-level concept-concept causal relationships, low-level gene-gene similarities, and concept-gene correspondences.
Therefore, it can be adapted to other causal analysis tasks where similar multi-level structure is available, as long as the data processing methods are adjusted to the corresponding data types.}

\subsection{Limitations and Future Work}

\myparagraph{Integration of multi-omics data}.
B2 noted that cellular behaviors are regulated by multiple interacting modalities, while the current causal graph models only one data modality at a time.
He suggested extending {\sys} to jointly analyze gene expression and spatial transcriptomics within a unified causal network.
However, effectively aligning heterogeneous data modalities and reliably analyzing cross-modal causal relationships remain challenging, making multi-omics causal analysis an important direction for future work.

\myparagraph{\pcheng{Scalability of dense source-target relationships}}\pcheng{.}
\pcheng{Although the \emph{Causal Graph Visualization} satisfies our requirements for jointly presenting causal relationships and gene-gene similarities, its readability may decrease when a concept is connected to a large number of concepts. In such cases, forcing all concepts to border the same concept may lead to crowded regions and make the left-to-right relationship harder to perceive. This limitation could be mitigated in future work by reducing the complexity, such as filtering weak causal links, aggregating low-impact concepts, or introducing hierarchical concept structure.}


\myparagraph{Usability evaluation}.
\pcheng{Currently, to allow B1 to focus more on analytical tasks rather than interface operation, we implemented the pair analytics protocol in the case study, which helped us discuss the analysis results.
However, when the proposed tool is deployed in real-world scenarios, experts will need to navigate the tool on their own.
Under such conditions, the learning curve remains unexplored, and the usability of complex interactions without guidance requires further validation.
In addition, the evaluation is also limited by the small number of participating experts.
Therefore, we plan to share {\sys} with a broader community of biologists and collect their feedback to iteratively improve its usability.}

\section{Conclusion}
In this paper, we propose an interactive visual analysis \pcheng{tool} to explore, validate, and refine causal relationships in virtual cells.
The core contribution is a gene-similarity-aware causal graph visualization \pcheng{that simultaneously preserves concept-concept causal relationships, concept-gene correspondences, and gene-gene similarities.}
By integrating causal path and counterfactual clustering visualizations, our system enables experts to interactively correct the causal graph and identify key regulatory genes of interest.
Case studies and domain expert feedback demonstrate that the proposed \pcheng{tool} effectively supports causal-driven biological research and helps uncover novel biological mechanisms.
\pcheng{Beyond virtual cells, this work may also be adapted to causal visual analysis tasks with the similar multi-level structure, where users need to inspect high-level causal relationships, low-level domain-specific similarities, and correspondences between these two levels.}


\acknowledgments{
The work is supported by Yuelushan Laboratory Breeding Program (Grant No. YLS-2025-ZY01015), the National Natural Science Foundation of China (Grant Nos. 62225205, 62532005, 62402167), the Science and Technology Program of Changsha (kh2301011), the Major Science and Technology Research Projects of Hunan Province (Grant Nos. 2024QK2010, 2024QK2009), the Yunnan Provincial Major Science and Technology Special Plan Projects (Grant No. 202502AD080009), the Yunnan Science and Technology Talents and Platforms Program (202605AK340003), Project of Yuelushan Center for Industrial Innovation (Grant No. 2025TCII0206), the Hunan Natural Science Foundation (Grant No. 2025JJ60419), and the Science and Technology Innovation Program of Hunan Province (Grant No. 2023ZJ1080).
}

\bibliographystyle{abbrv-doi-hyperref}

\bibliography{reference}

@article{subramanian2005gene,
  title={Gene set enrichment analysis: a knowledge-based approach for interpreting genome-wide expression profiles},
  author={Subramanian, Aravind and Tamayo, Pablo and Mootha, Vamsi K and Mukherjee, Sayan and Ebert, Benjamin L and Gillette, Michael A and Paulovich, Amanda and Pomeroy, Scott L and Golub, Todd R and Lander, Eric S and others},
  journal={Proceedings of the national academy of sciences},
  volume={102},
  number={43},
  pages={15545--15550},
  year={2005},
  publisher={National Academy of Sciences},
  doi={10.1073/pnas.0506580102}
}

@inproceedings{satopaa2011finding,
  title={Finding a" kneedle" in a haystack: Detecting knee points in system behavior},
  author={Satopaa, Ville and Albrecht, Jeannie and Irwin, David and Raghavan, Barath},
  booktitle={2011 31st international conference on distributed computing systems workshops},
  pages={166--171},
  year={2011},
  organization={IEEE},
  doi={10.1109/ICDCSW.2011.20}
}

@article{friedman1937use,
  title={The use of ranks to avoid the assumption of normality implicit in the analysis of variance},
  author={Friedman, Milton},
  journal={Journal of the american statistical association},
  volume={32},
  number={200},
  pages={675--701},
  year={1937},
  publisher={Taylor \& Francis},
  doi={10.2307/2279372}
}

@article{chen2025interactive,
  title={Interactive Hybrid Rice Breeding with Parametric Dual Projection},
  author={Chen, Changjian and Wang, Pengcheng and Lyu, Fei and Tang, Zhuo and Yang, Li and Wang, Long and Cai, Yong and Yu, Feng and Li, Kenli},
  journal={IEEE Transactions on Visualization and Computer Graphics},
  year={2025},
  publisher={IEEE},
  doi={10.1109/TVCG.2025.3634640}
}

@article{scholkopf2021toward,
  title={Toward causal representation learning},
  author={Sch{\"o}lkopf, Bernhard and Locatello, Francesco and Bauer, Stefan and Ke, Nan Rosemary and Kalchbrenner, Nal and Goyal, Anirudh and Bengio, Yoshua},
  journal={Proceedings of the IEEE},
  volume={109},
  number={5},
  pages={612--634},
  year={2021},
  doi={10.1109/JPROC.2021.3058954},
  publisher={IEEE}
}

@article{burgess2017transcriptional,
  title={Transcriptional control of photosynthetic capacity: conservation and divergence from Arabidopsis to rice},
  author={Burgess, Steven J and Reyna-Llorens, Ivan and Stevenson, Susan R and Bowman, Pallavi N and Townsend, Alison J and Kelly, Steven},
  journal={New Phytologist},
  volume={216},
  number={2},
  pages={510--528},
  year={2017},
  doi={10.1111/nph.14682},
  publisher={Wiley Online Library}
}

@article{zeeman2010starch,
  title={Starch: its metabolism, evolution, and biotechnological modification in plants},
  author={Zeeman, Samuel C and Kossmann, Jens and Smith, Alison M},
  journal={Annual review of plant biology},
  volume={61},
  pages={209--234},
  year={2010},
  doi={10.1146/annurev-arplant-042809-112301},
  publisher={Annual Reviews}
}

@article{yamakawa2007comprehensive,
  title={Comprehensive expression profiling of rice grain filling-related genes under high temperature using DNA microarray},
  author={Yamakawa, Hiromoto and Hirose, Tatsuro and Kuroda, Masaharu and Yamaguchi, Takeshi},
  journal={Plant physiology},
  volume={144},
  number={1},
  pages={258--277},
  year={2007},
  doi={10.1104/pp.107.098665},
  publisher={American Society of Plant Biologists}
}

@article{sakai2013rice,
  title={The Rice Annotation Project Database (RAP-DB): an integrative hub for rice genomics},
  author={Sakai, Hiroaki and Lee, Steven S and Tanaka, Tsuyoshi and Numa, Hisataka and Itoh, Takeshi and others},
  journal={Nucleic Acids Research},
  volume={41},
  number={D1},
  pages={D1196--D1205},
  year={2013},
  doi={10.1093/nar/gkj094},
  publisher={Oxford University Press}
}

@article{kawahara2013improvement,
  title={Improvement of the Oryza sativa Nipponbare reference genome sequence and annotation: a report from the Rice Annotation Project},
  author={Kawahara, Yoshihiro and de la Bastide, Marc and Hamilton, John P and Kanamori, Hiroyuki and McCombie, W Richard and Ouyang, Shu and others},
  journal={Rice},
  volume={6},
  pages={1--10},
  year={2013},
  doi={10.1186/1939-8433-6-4},
  publisher={Springer}
}

@article{aurenhammer1991voronoi,
  title={Voronoi diagrams—a survey of a fundamental geometric data structure},
  author={Aurenhammer, Franz},
  journal={ACM computing surveys (CSUR)},
  volume={23},
  number={3},
  pages={345--405},
  year={1991},
  doi={10.1145/116873.116880},
  publisher={ACM New York, NY, USA}
}

@inproceedings{macqueen1967some,
  title={Some methods for classification and analysis of multivariate observations},
  author={MacQueen, James and others},
  booktitle={Proceedings of the fifth Berkeley symposium on mathematical statistics and probability},
  volume={1},
  number={14},
  pages={281--297},
  year={1967},
  organization={Oakland, CA, USA}
}

@article{maglott2005entrez,
  title={Entrez Gene: gene-centered information at NCBI},
  author={Maglott, Donna and Ostell, Jim and Pruitt, Kim D and Tatusova, Tatiana},
  journal={Nucleic acids research},
  volume={33},
  number={suppl\_1},
  pages={D54--D58},
  year={2005},
  doi={10.1093/nar/gkl993},
  publisher={Oxford University Press}
}

@article{jan2022retrograde,
  title={Retrograde and anterograde signaling in the crosstalk between chloroplast and nucleus},
  author={Jan, Masood and Liu, Zhixin and Rochaix, Jean-David and Sun, Xuwu},
  journal={Frontiers in Plant Science},
  volume={13},
  pages={980237},
  year={2022},
  doi={10.3389/fpls.2022.980237},
  publisher={Frontiers Media SA}
}

@article{ruscone2025neko,
  title={NeKo: a tool for automatic network construction from prior knowledge},
  author={Ruscone, Marco and Tsirvouli, Eirini and Checcoli, Andrea and Turei, Denes and Barillot, Emmanuel and Saez-Rodriguez, Julio and Martignetti, Loredana and Flobak, {\AA}smund and Calzone, Laurence},
  journal={PLOS Computational Biology},
  volume={21},
  number={9},
  pages={e1013300},
  year={2025},
  doi={10.1371/journal.pcbi.1013300},
  publisher={Public Library of Science San Francisco, CA USA}
}

@inproceedings{gansner2010gmap,
  title={GMap: Visualizing graphs and clusters as maps},
  author={Gansner, Emden R and Hu, Yifan and Kobourov, Stephen},
  booktitle={2010 IEEE Pacific visualization symposium (PacificVis)},
  pages={201--208},
  year={2010},
  doi={10.1109/PACIFICVIS.2010.5429590},
  organization={IEEE}
}

@article{aibar2017scenic,
  title={SCENIC: single-cell regulatory network inference and clustering},
  author={Aibar, Sara and Gonz{\'a}lez-Blas, Carmen Bravo and Moerman, Thomas and Huynh-Thu, V{\^a}n Anh and Imrichova, Hana and Hulselmans, Gert and Rambow, Florian and Marine, Jean-Christophe and Geurts, Pierre and Aerts, Jan and others},
  journal={Nature methods},
  volume={14},
  number={11},
  pages={1083--1086},
  year={2017},
  doi={10.1038/nmeth.4463},
  publisher={Nature Publishing Group US New York}
}

@article{liu2024cwgcna,
  title={{CWGCNA}: an R package to perform causal inference from the {WGCNA} framework},
  author={Liu, Yu},
  journal={NAR Genomics and Bioinformatics},
  volume={6},
  number={2},
  pages={lqae042},
  year={2024},
  doi={10.1093/nargab/lqae042},
  publisher={Oxford University Press}
}

@article{wang2026osmads18,
  title={The OsMADS18-OsbZIP60 module plays a critical role in influencing grain chalkiness in rice},
  author={Wang, Xiaohang and Liu, Wenxin and Xing, Jianying and Xue, Shangyong and Zhou, Dongxiao and Zhou, Ganghua and Xu, Wenhui and Li, Zhe and Liu, Yutong and Yun, Dae-Jin and Xu, Zheng-Yi},
  journal={Science China Life Sciences},
  volume={69},
  pages={651--661},
  year={2026},
  doi={10.1007/s11427-025-3129-3},
  url={https://doi.org/10.1007/s11427-025-3129-3}
}

@inproceedings{arias2011pair,
  title={Pair analytics: Capturing reasoning processes in collaborative visual analytics},
  author={Arias-Hernandez, Richard and Kaastra, Linda T and Green, Tera M and Fisher, Brian},
  booktitle={Proceedings of the IEEE International Conference on System Sciences},
  pages={1--10},
  year={2011},
  doi={10.1109/HICSS.2011.339}
}

@article{chen2025human,
  author={Chen, Changjian and Lv, Fei and Guan, Yalong and Wang, Pengcheng and Yu, Shengjie and Zhang, Yifan and Tang, Zhuo},
  journal={IEEE Transactions on Visualization and Computer Graphics}, 
  title={Human-Guided Image Generation for Expanding Small-Scale Training Image Datasets}, 
  year={2025},
  volume={31},
  number={6},
  pages={3809-3821},
  doi={10.1109/TVCG.2025.3567053}
}

@book{spirtes2000causation,
  title={Causation, prediction, and search},
  author={Spirtes, Peter and Glymour, Clark N and Scheines, Richard},
  year={2000},
  publisher={MIT press}
}

@article{ashburner2000gene,
  title={Gene ontology: tool for the unification of biology},
  author={Ashburner, Michael and Ball, Catherine A and Blake, Judith A and Botstein, David and Butler, Heather and Cherry, J Michael and Davis, Allan P and Dolinski, Kara and Dwight, Selina S and Eppig, Janan T and others},
  journal={Nature genetics},
  volume={25},
  number={1},
  pages={25--29},
  year={2000},
  doi={10.1038/75556},
  publisher={Nature Publishing Group}
}

@article{TANG202448,
title = {ArtEyer: Enriching GPT-based agents with contextual data visualizations for fine art authentication},
journal = {Visual Informatics},
volume = {8},
number = {4},
pages = {48-59},
year = {2024},
issn = {2468-502X},
doi = {https://doi.org/10.1016/j.visinf.2024.11.001},
url = {https://www.sciencedirect.com/science/article/pii/S2468502X24000664},
author = {Tan Tang and Yanhong Wu and Junming Gao and Kejia Ruan and Yanjie Zhang and Shuainan Ye and Yingcai Wu and Xiaojiao Chen}
}

@article{PENG2025100234,
title = {Contextualized visual analytics for multivariate events},
journal = {Visual Informatics},
volume = {9},
number = {2},
pages = {100234},
year = {2025},
issn = {2468-502X},
doi = {https://doi.org/10.1016/j.visinf.2025.100234},
url = {https://www.sciencedirect.com/science/article/pii/S2468502X25000099},
author = {Lei Peng and Ziyue Lin and Natalia Andrienko and Gennady Andrienko and Siming Chen}
}

@article{wold1987pca,
  title={Principal component analysis},
  author={Wold, Svante and Esbensen, Kim and Geladi, Paul},
  journal={Chemometrics and Intelligent Laboratory Systems},
  volume={2},
  number={1-3},
  pages={37--52},
  year={1987},
  doi={10.1007/springerreference_84147}
}

@article{van2008tsne,
  title={Visualizing data using t-SNE.},
  author={Van der Maaten, Laurens and Hinton, Geoffrey},
  journal={Journal of Machine Learning Research},
  volume={9},
  number={11},
  pages={2579--2605},
  year={2008}
}

@article{dennig2023fs,
  title={{FS/DS}: A theoretical framework for the dual analysis of feature space and data space},
  author={Dennig, Frederik L and Miller, Matthias and Keim, Daniel A and El-Assady, Mennatallah},
  journal={IEEE Transactions on Visualization and Computer Graphics},
  year={2023},
  doi={10.1109/TVCG.2023.3288356}
}

@article{cowen2017network,
  title={Network propagation: a universal amplifier of genetic associations},
  author={Cowen, Lenore and Ideker, Trey and Raphael, Benjamin J and Sharan, Roded},
  journal={Nature Reviews Genetics},
  volume={18},
  number={9},
  pages={551--562},
  year={2017},
  publisher={Nature Publishing Group}
}

@article{regev2017human,
  title={The human cell atlas},
  author={Regev, Aviv and Teichmann, Sarah A and Lander, Eric S and Amit, Ido and Benoist, Christophe and Birney, Ewan and Bodenmiller, Bernd and Campbell, Peter and Carninci, Piero and Clatworthy, Menna and others},
  journal={elife},
  volume={6},
  pages={e27041},
  year={2017},
  doi = {10.7554/eLife.27041},
  publisher={eLife Sciences Publications, Ltd}
}

@article{hoel2013quantifying,
  title={Quantifying causal emergence shows that macro can beat micro},
  author={Hoel, Erik P and Albantakis, Larissa and Tononi, Giulio},
  journal={Proceedings of the National Academy of Sciences},
  volume={110},
  number={49},
  pages={19790--19795},
  year={2013},
  doi = {10.1073/pnas.1314922110},
  publisher={National Academy of Sciences}
}

@article{langfelder2008wgcna,
  title={WGCNA: an R package for weighted correlation network analysis},
  author={Langfelder, Peter and Horvath, Steve},
  journal={BMC bioinformatics},
  volume={9},
  number={1},
  pages={559},
  year={2008},
  doi = {10.1186/1471-2105-9-559},
  publisher={Springer}
}

@article{wang2025single,
  title={A single-cell multi-omics atlas of rice},
  author={Wang, Xiangyu and Huang, Huanwei and Jiang, Sanjie and Kang, Jingmin and Li, Dongwei and Wang, Kailai and Xie, Shang and Tong, Cheng and Liu, Chaofan and Hu, Guihua and others},
  journal={Nature},
  volume={644},
  number={8077},
  pages={722--730},
  year={2025},
  doi = {10.1038/s41586-025-09251-0},
  publisher={Nature Publishing Group UK London}
}

@article{mazzarello1999unifying,
  title={A unifying concept: the history of cell theory},
  author={Mazzarello, Paolo},
  journal={Nature cell biology},
  volume={1},
  number={1},
  pages={E13--E15},
  year={1999},
  doi = {10.1038/8964},
  publisher={Nature Publishing Group}
}

@article{rood2025human,
  title={The Human Cell Atlas from a cell census to a unified foundation model},
  author={Rood, Jennifer E and Wynne, Samantha and Robson, Lucia and Hupalowska, Anna and Randell, John and Teichmann, Sarah A and Regev, Aviv},
  journal={Nature},
  volume={637},
  number={8048},
  pages={1065--1071},
  year={2025},
  doi = {10.1038/s41586-024-08338-4},
  publisher={Nature Publishing Group UK London}
}

@article{gao2025causal,
  title={Causal disentanglement for single-cell representations and controllable counterfactual generation},
  author={Gao, Yicheng and Dong, Kejing and Shan, Caihua and Li, Dongsheng and Liu, Qi},
  journal={Nature communications},
  volume={16},
  number={1},
  pages={6775},
  year={2025},
  doi = {10.1038/s41467-025-62008-1},
  publisher={Nature Publishing Group UK London}
}

@article{zhao2023large,
  title={Large-scale cell representation learning via divide-and-conquer contrastive learning},
  author={Zhao, Suyuan and Zhang, Jiahuan and Nie, Zaiqing},
  journal={arXiv preprint arXiv:2306.04371},
  doi = {10.48550/arXiv.2306.04371},
  year={2023}
}

@inproceedings{lopez2023learning,
  title={Learning causal representations of single cells via sparse mechanism shift modeling},
  author={Lopez, Romain and Tagasovska, Natasa and Ra, Stephen and Cho, Kyunghyun and Pritchard, Jonathan and Regev, Aviv},
  booktitle={Conference on Causal Learning and Reasoning},
  pages={662--691},
  year={2023},
  doi={10.48550/arXiv.2211.03553},
  organization={PMLR}
}

@article{lotfollahi2023predicting,
  title={Predicting cellular responses to complex perturbations in high-throughput screens},
  author={Lotfollahi, Mohammad and Klimovskaia Susmelj, Anna and De Donno, Carlo and Hetzel, Leon and Ji, Yuge and Ibarra, Ignacio L and Srivatsan, Sanjay R and Naghipourfar, Mohsen and Daza, Riza M and Martin, Beth and others},
  journal={Molecular systems biology},
  volume={19},
  number={6},
  pages={MSB202211517},
  year={2023},
  doi = {10.15252/msb.202211517},
  publisher={Springer}
}

@article{bereket2023modelling,
  title={Modelling cellular perturbations with the sparse additive mechanism shift variational autoencoder},
  author={Bereket, Michael and Karaletsos, Theofanis},
  journal={Advances in Neural Information Processing Systems},
  volume={36},
  pages={1--12},
  year={2023}
}

@article{bunne2024build,
  title={How to build the virtual cell with artificial intelligence: Priorities and opportunities},
  author={Bunne, Charlotte and Roohani, Yusuf and Rosen, Yanay and Gupta, Ankit and Zhang, Xikun and Roed, Marcel and Alexandrov, Theo and AlQuraishi, Mohammed and Brennan, Patricia and Burkhardt, Daniel B and others},
  journal={Cell},
  volume={187},
  number={25},
  pages={7045--7063},
  year={2024},
  doi = {10.1016/j.cell.2024.11.015},
  publisher={Elsevier}
}

@article{yang2024genecompass,
  title={GeneCompass: deciphering universal gene regulatory mechanisms with a knowledge-informed cross-species foundation model},
  author={Yang, Xiaodong and Liu, Guole and Feng, Guihai and Bu, Dechao and Wang, Pengfei and Jiang, Jie and Chen, Shubai and Yang, Qinmeng and Miao, Hefan and Zhang, Yiyang and others},
  journal={Cell Research},
  volume={34},
  number={12},
  pages={830--845},
  year={2024},
  doi = {10.1038/s41422-024-01034-y},
  publisher={Springer Nature Singapore Singapore}
}

@article{zhang2024scdisinfact,
  title={scDisInFact: disentangled learning for integration and prediction of multi-batch multi-condition single-cell RNA-sequencing data},
  author={Zhang, Ziqi and Zhao, Xinye and Bindra, Mehak and Qiu, Peng and Zhang, Xiuwei},
  journal={Nature Communications},
  volume={15},
  number={1},
  pages={912},
  year={2024},
  doi = {10.1038/s41467-024-45227-w},
  publisher={Nature Publishing Group UK London}
}

@article{theodoris2023transfer,
  title={Transfer learning enables predictions in network biology},
  author={Theodoris, Christina V and Xiao, Ling and Chopra, Anant and Chaffin, Mark D and Al Sayed, Zeina R and Hill, Matthew C and Mantineo, Helene and Brydon, Elizabeth M and Zeng, Zexian and Liu, X Shirley and others},
  journal={Nature},
  volume={618},
  number={7965},
  pages={616--624},
  year={2023},
  doi = {10.1038/s41586-023-06139-9},
  publisher={Nature Publishing Group UK London}
}

@article{roohani2024predicting,
  title={Predicting transcriptional outcomes of novel multigene perturbations with GEARS},
  author={Roohani, Yusuf and Huang, Kexin and Leskovec, Jure},
  journal={Nature Biotechnology},
  volume={42},
  number={6},
  pages={927--935},
  year={2024},
  doi = {10.1038/s41587-023-01905-6},
  publisher={Nature Publishing Group US New York}
}

@article{piran2024disentanglement,
  title={Disentanglement of single-cell data with biolord},
  author={Piran, Zoe and Cohen, Niv and Hoshen, Yedid and Nitzan, Mor},
  journal={Nature Biotechnology},
  volume={42},
  number={11},
  pages={1678--1683},
  year={2024},
  doi = {10.1038/s41587-023-02079-x},
  publisher={Nature Publishing Group US New York}
}

@article{lopez2018deep,
  title={Deep generative modeling for single-cell transcriptomics},
  author={Lopez, Romain and Regier, Jeffrey and Cole, Michael B and Jordan, Michael I and Yosef, Nir},
  journal={Nature methods},
  volume={15},
  number={12},
  pages={1053--1058},
  year={2018},
  doi = {10.1038/s41592-018-0229-2},
  publisher={Nature Publishing Group US New York}
}

@article{lotfollahi2019scgen,
  title={scGen predicts single-cell perturbation responses},
  author={Lotfollahi, Mohammad and Wolf, F Alexander and Theis, Fabian J},
  journal={Nature methods},
  volume={16},
  number={8},
  pages={715--721},
  year={2019},
  doi = {10.1038/s41592-019-0494-8},
  publisher={Nature Publishing Group US New York}
}

@article{cui2024scgpt,
  title={scGPT: toward building a foundation model for single-cell multi-omics using generative AI},
  author={Cui, Haotian and Wang, Chloe and Maan, Hassaan and Pang, Kuan and Luo, Fengning and Duan, Nan and Wang, Bo},
  journal={Nature methods},
  volume={21},
  number={8},
  pages={1470--1480},
  year={2024},
  doi = {10.1038/s41592-024-02201-0},
  publisher={Nature Publishing Group US New York}
}

@article{teng2023vispur,
  title={VISPUR: Visual Aids for Identifying and Interpreting Spurious Associations in Data-Driven Decisions},
  author={Teng, Xian and Ahn, Yongsu and Lin, Yu-Ru},
  journal={IEEE Transactions on Visualization and Computer Graphics},
  volume={30},
  number={1},
  pages={219--229},
  year={2023},
  doi = {10.1109/TVCG.2023.3326587},
  publisher={IEEE}
}

@article{weinberg2025causality,
  title={Causality from bottom to top: A survey},
  author={Weinberg, Abraham Itzhak and Premebida, Cristiano and Faria, Diego Resende},
  journal={Machine Learning},
  volume={114},
  number={11},
  pages={234},
  year={2025},
  doi = {10.1007/s10994-025-06855-5},
  publisher={Springer}
}

@article{wang2015visual,
  title={The visual causality analyst: An interactive interface for causal reasoning},
  author={Wang, Jun and Mueller, Klaus},
  journal={IEEE transactions on visualization and computer graphics},
  volume={22},
  number={1},
  pages={230--239},
  year={2015},
  doi = {10.1109/TVCG.2015.2467931},
  publisher={IEEE}
}

@inproceedings{bae2017understanding,
  title={Understanding Indirect Causal Relationships in Node-Link Graphs},
  author={Bae, Juhee and Helldin, Tove and Riveiro, Maria},
  booktitle={Computer graphics forum},
  volume={36},
  number={3},
  pages={411--421},
  year={2017},
  doi = {10.1111/cgf.13198},
  organization={Wiley Online Library}
}

@article{forbes2017dynamic,
  title={Dynamic influence networks for rule-based models},
  author={Forbes, Angus G and Burks, Andrew and Lee, Kristine and Li, Xing and Boutillier, Pierre and Krivine, Jean and Fontana, Walter},
  journal={IEEE transactions on visualization and computer graphics},
  volume={24},
  number={1},
  pages={184--194},
  year={2017},
  doi = {10.1109/TVCG.2017.2745280},
  publisher={IEEE}
}

@inproceedings{yan2020silva,
  title={Silva: Interactively assessing machine learning fairness using causality},
  author={Yan, Jing Nathan and Gu, Ziwei and Lin, Hubert and Rzeszotarski, Jeffrey M},
  booktitle={Proceedings of the 2020 chi conference on human factors in computing systems},
  pages={1--13},
  doi = {10.1145/3313831.3376447},
  year={2020}
}

@article{xie2020visual,
  title={A visual analytics approach for exploratory causal analysis: Exploration, validation, and applications},
  author={Xie, Xiao and Du, Fan and Wu, Yingcai},
  journal={IEEE Transactions on Visualization and Computer Graphics},
  volume={27},
  number={2},
  pages={1448--1458},
  year={2020},
  doi = {10.1109/TVCG.2020.3028957},
  publisher={IEEE}
}

@article{hoque2021outcome,
  title={Outcome-explorer: A causality guided interactive visual interface for interpretable algorithmic decision making},
  author={Hoque, Md Naimul and Mueller, Klaus},
  journal={IEEE Transactions on Visualization and Computer Graphics},
  volume={28},
  number={12},
  pages={4728--4740},
  year={2021},
  doi = {10.1109/TVCG.2021.3102051},
  publisher={IEEE}
}

@article{kaul2021improving,
  title={Improving visualization interpretation using counterfactuals},
  author={Kaul, Smiti and Borland, David and Cao, Nan and Gotz, David},
  journal={IEEE Transactions on Visualization and Computer Graphics},
  volume={28},
  number={1},
  pages={998--1008},
  year={2021},
  doi = {10.1109/TVCG.2021.3114779},
  publisher={IEEE}
}

@article{ghai2022d,
  title={D-BIAS: A causality-based human-in-the-loop system for tackling algorithmic bias},
  author={Ghai, Bhavya and Mueller, Klaus},
  journal={IEEE Transactions on Visualization and Computer Graphics},
  volume={29},
  number={1},
  pages={473--482},
  year={2022},
  doi = {10.1109/TVCG.2022.3209484},
  publisher={IEEE}
}

@inproceedings{guo2023causalvis,
  title={Causalvis: Visualizations for causal inference},
  author={Guo, Grace and Karavani, Ehud and Endert, Alex and Kwon, Bum Chul},
  booktitle={Proceedings of the 2023 CHI conference on human factors in computing systems},
  pages={1--20},
  doi = {10.1145/3544548.3581236},
  year={2023}
}

@article{li2025causality,
  title={Causality-based Visual Analytics of Sentiment Contagion in Social Media Topics},
  author={Li, Renzhong and Ye, Shuainan and Lin, Yuchen and Zhou, Buwei and Kang, Zhining and Peng, Tai-Quan and Fu, Wenhao and Tang, Tan and Wu, Yingcai},
  journal={IEEE Transactions on Visualization and Computer Graphics},
  year={2025},
  doi = {10.1109/TVCG.2025.3633839},
  publisher={IEEE}
}

@article{fan2024visual,
  title={Visual analysis of multi-outcome causal graphs},
  author={Fan, Mengjie and Yu, Jinlu and Weiskopf, Daniel and Cao, Nan and Wang, Huai-Yu and Zhou, Liang},
  journal={IEEE Transactions on Visualization and Computer Graphics},
  volume={31},
  number={1},
  pages={656--666},
  year={2024},
  doi = {10.1109/TVCG.2024.3456346},
  publisher={IEEE}
}

@article{borland2024using,
  title={Using counterfactuals to improve causal inferences from visualizations},
  author={Borland, David and Wang, Arran Zeyu and Gotz, David},
  journal={IEEE Computer Graphics and Applications},
  volume={44},
  number={1},
  pages={95--104},
  year={2024},
  doi = {10.1109/MCG.2023.3338788},
  publisher={IEEE}
}

@article{zhang2025causalchat,
  title={Causalchat: Interactive causal model development and refinement using large language models},
  author={Zhang, Yanming and Kota, Akshith and Papenhausen, Eric and Mueller, Klaus},
  journal={IEEE Transactions on Visualization and Computer Graphics},
  year={2025},
  doi = {10.1109/TVCG.2025.3602448},
  publisher={IEEE}
}

@inproceedings{vo2020visual,
  title={Visual causality: Investigating graph layouts for understanding causal processes},
  author={Vo, Dong-Bach and Lazarova, Kristina and Purchase, Helen C and McCann, Mark},
  booktitle={International Conference on Theory and Application of Diagrams},
  pages={332--347},
  year={2020},
  doi = {10.1007/978-3-030-54249-8_26},
  organization={Springer}
}

@article{meng2023class,
  title={Class-constrained t-sne: combining data features and class probabilities},
  author={Meng, Linhao and van den Elzen, Stef and Pezzotti, Nicola and Vilanova, Anna},
  journal={IEEE Transactions on Visualization and Computer Graphics},
  volume={30},
  number={1},
  pages={164--174},
  year={2023},
  doi = {10.1109/TVCG.2023.3326600},
  publisher={IEEE}
}

\end{document}